\documentclass[11pt]{article}
\usepackage{verbatim}
\usepackage{epsfig}
\usepackage{amssymb,amsmath,amsthm}
\usepackage{relsize}
\usepackage{mathrsfs}
\usepackage{xcolor}
\usepackage{eucal}
\usepackage{epsfig}
\usepackage{epstopdf}
\usepackage{xchicago}

\newcommand {\citeAY} [1] {\citeNP {#1}}%
\newcommand {\citeAPY}[1] {\citeN  {#1}}%
\newcommand {\citePAY}[1] {\cite   {#1}}%
\renewcommand {\showoriginalref}[1]{}
\renewcommand {\showCODEN}[1]{}
\renewcommand {\showISSN}[1]{}
\renewcommand {\showMR}[3]{}

\newcommand\sect[1] {\ref{sec:#1}}

\newcommand\labsect[1] {\label{sec:#1}}

\begin{document}

\title{\bf Bounds for the response of viscoelastic composites under antiplane loadings in the time domain}
\author{Ornella Mattei $^{*}$ and Graeme W. Milton$^{**}$} 
\date{$^{*}$\small{DICATAM, University of Brescia, Via Branze, 43-25123, Brescia, Italy}\\
$^{**}$\small{Department of Mathematics, University of Utah, Salt Lake City, UT 84112, USA}}

\maketitle
\begin{abstract}
\noindent
In order to derive bounds on the strain and stress response of a two-comp\-o\-nent composite material with viscoelastic phases, we revisit the so-called analytic method \citePAY{Bergman:1978:DCC}, which allows one to approximate the complex effective tensor, function of the ratio of the component shear moduli, as the sum of poles weighted by positive semidefinite residue matrices. The novelty of the present investigation lies in the application of such a method, previously applied
(\citeAY{Milton:1980:BCD}; \citeAY{Bergman:1980:ESM}), to problems involving cyclic loadings in the frequency domain, to derive bounds in the \textit{time domain} for the antiplane viscoelasticity case. 

The position of the poles and the residues matrices are the variational parameters of the problem: the aim is to determine such parameters in order to have the minimum (or maximum) response at any given moment in time. All the information about the composite, such as the knowledge of the volume fractions or the transverse isotropy of the composite, is translated for each fixed pole configuration into (linear) constraints on the residues, the so-called sum rules. Further constraints can be obtained from the knowledge of the response of the composite at specific times (in this paper, for instance, we show how one can include information about the instantaneous and the long-term response of the composite). 

The linearity of the constraints, along with the observation that the response at a fixed time is linear in the residues, enables one to use the theory of linear programming to reduce the problem to one involving relatively few non-zero residues. Finally, bounds on the response are obtained by numerically optimizing over the pole positions. In the examples studied, the results turn out to be very accurate estimates: if sufficient information about the composite is available, the bounds can be quite tight over the entire range of time, allowing one to predict the transient behavior of the composite. 
Furthermore, the bounds incorporating the volume fractions (and possibly transverse isotropy) can be extremely tight at certain specific times: thus measuring the response at such times, and using the bounds in an inverse fashion, gives very tight bounds on the volume fraction of the phases in the composite. 
\end{abstract}

\section{Introduction}\label{HIntroduction}
\setcounter{equation}{0}

The problem of calculating the mechanical response of a composite material has been extensively investigated in the literature, with particular attention being paid to the derivation of approximation
formulae and bounds  on the effective properties of the composite. This has taken precedence over the determination of the exact response of the material, which represents a difficult task even in the rare situations where the microstructure is known.

Historically, in the elasticity case, the determination of bounds on the overall properties of the composite followed from the formulation of suitable extremum variational principles, as illustrated, for instance, in the pioneering work of \citeAPY{Hill:1952:EBC}%
\index{Hill bounds}
and  Hashin and Shtrikman (\citeyearNP{Hashin:1962:VAT},\citeyearNP{Hashin:1963:VAT}),%
\index{Hashin-Shtrikman bounds}
which paved the way to the calculation of rigorous geometry-independent bounds. Such bounds have proven to be useful benchmarks for testing experimental results and for setting limits on the range of possible responses, which is relevant when one is optimizing the microstructure to maximize performance. Variational principles%
\index{variational principles}
are useful even when some of the moduli are negative:%
\index{negative moduli}
in \citeAY{Kochmann:2015:RBE}, the authors have ruled out the possibility of achieving very stiff statically stable composites by combining materials with positive and negative moduli, as suggested by \citeAPY{Lakes:2002:DSE}. The variational method is especially powerful when coupled with the translation method%
\index{translation method}
of Tartar and Murat, and Lurie and Cherkaev [see Chapter 24 of \citeAY{Milton:2002:TOC} for relevant references], who used it to derive optimal bounds on the possible effective conductivity tensors of two and three dimensional two-phase conducting composites.%
\index{conducting composites}
The translation method can also be used to bound the response of inhomogeneous bodies or, inversely, to bound the volume fraction%
\index{bounding!volume fraction}
of the phases from measurements of the fields at the surface of the body \cite{Kang:2013:BVF3d}. For surveys of bounds on the effective properties of composites (and the various methods used to derive them) see the books of \citeAPY{Cherkaev:2000:VMS}, \citeAPY{Milton:2002:TOC}, \citeAPY{Allaire:2002:SOH}, \citeAPY{Torquato:2002:RHM}, \citeAPY{Tartar:2009:GTH} and references therein. 

For the viscoelasticity%
\index{viscoelasticity}
case, instead, the lack in the time domain of variational formulations analogous to the ones for the elasticity problem first 
prompted several authors to apply the correspondence principle%
\index{correspondence principle}
(\citeAY{Hashin:1965:VBH}; \citeAY{Christensen:1971:TVI}) to the well-established results of the elastic problem 
in order to study the linear viscoelastic response of composites subject to a cyclic loading with a certain frequency. In fact, for low frequency harmonic vibrations, where the inertia effects can be neglected and the viscoelastic loss is small compared to the elastic moduli (thus ruling out phases which are viscous fluids or gel-like), the bounds that have been obtained which couple the effective properties with the derivatives of the effective properties with respect to the moduli (such as those obtained by \citeAPY{Prager:1969:IVB}) when the moduli are real, imply the correspondence principle bounds on the
complex effective properties \citePAY{Schulgasser:1976:BEP}.  The correspondence principle itself requires justification, and this justification is provided by the analyticity%
\index{analyticity as functions of the component moduli}
of the effective moduli as functions of the component moduli [see Section 11.4 in \citeAY{Milton:2002:TOC}].
This analyticity was first recognized by \citeAPY{Bergman:1978:DCC} in the context of the dielectric problem%
\index{dielectric problem}
for composites of two isotropic components. Some of the assumptions underlying his initial analysis were incorrect \citePAY{Milton:1979:TST}: in particular, he assumed that for periodic media the effective dielectric constant is a rational function of the component moduli. This is not true in checkerboard geometries,%
\index{checkerboard geometries}
where the function has a branch cut,%
\index{branch cut}
and if branch cuts can appear one may ask: why cannot they occur when the dielectric constants have positive imaginary parts, and not just when the ratio of the dielectric constants is real and negative?

A plausible justification for Bergman's approach was first provided by \citeAPY{Milton:1981:BCP}, based on the assumption that the composite could be approximated by a large network containing two types of impedances,%
\index{resistor network}
where the length scale of the network grid is much smaller than that of the composite microstructure. Later, Golden and Papanicolaou
(\citeyearNP{Golden:1983:BEP}, \citeyearNP{Golden:1985:BEP}) gave a rigorous proof of the analytic properties and,
moreover, they established the analyticity for multicomponent media and obtained a representation formula%
\index{representation formula}
for the effective conductivity tensor as a function of the component conductivities which separates the
dependence on the component conductivities (contained in an appropriate integral kernel) from the dependence on the microstructure (the relevant information about which is contained in a positive measure).
These analytic properties enabled Milton (\citeyearNP{Milton:1980:BCD}, \citeyearNP{Milton:1981:BCP}, \citeyearNP{Milton:1981:BTO}) and 
Bergman (\citeyearNP{Bergman:1980:ESM}, \citeyearNP{Bergman:1982:RBC}), in independent works, to show that the complex 
effective dielectric constant (no matter how lossy the materials are, provided only that the quasistatic approximation is valid) is confined to a nested set of lens shaped regions%
\index{nested lens shaped bounds}
in the complex plane, where the relevant lens shaped region is determined by what information is known about the composite (such as the volume fractions of the phases, whether it is isotropic or transversely isotropic, the values of real or complex dielectric constant at a set of other frequencies). Some, but not all, of these bounds are implied by bounds on Stieltjes functions:%
\index{Stieltjes functions}
see the discussion in the Introduction of \citeAY{Milton:1987:MCEb} and references therein. 
With a small modification the bounds in \citeAY{Milton:1981:BTO} also apply to the related problem of bounding the viscoelastic moduli%
\index{bounds!viscoelastic moduli of homogeneous materials}
of homogeneous materials at one frequency, 
given the viscoelastic moduli at several other frequencies \citePAY{Eyre:2000:BIC}.

The bounds in the two-dimensional case immediately imply bounds for the mathematically equivalent problem of antiplane elasticity%
\index{antiplane elasticity}
(in this connnection it is to be noted 
that the claim of \citeAPY{Bergman:1980:ESM} that the two-dimensional bounds of \citeAPY{Milton:1980:BCD} were not attained by assemblages of doubly coated cylinders%
\index{assemblages!doubly coated cylinders}
was, in fact, wrong: curiously an earlier version of his paper, which did not reference the doubly coated cylinder geometry in \citeAPY{Milton:1980:BCD}, but which did reference the paper, had claimed that the three-dimensional bounds were attained by doubly coated spheres, which is incorrect). 
Interestingly in the two dimensional case (i.e, the antiplane elastic or antiplane viscoelastic case) for two component media (and polycrystals%
\index{polycrystals}
of a single crystal), the characterization of the analytic properties is complete, and moreover the functional dependence of the matrix valued effective dielectric tensor on the component moduli (or on the crystal tensor) can be mimicked to an arbitrarily high degree of approximation by a hierarchical laminate structure%
\index{laminates!hierarchical}
(\citeAY{Milton:1986:APLG}; \citeAY{Clark:1994:MEC}) [see also Section 18.5 in \citeAY{Milton:2002:TOC}] or when the two-component composite is isotropic by an assemblage of multicoated cylinders%
\index{assemblage!multicoated cylinder}
\citePAY{Milton:1981:BCP} [see the paragraph preceding Section VI]. Consequently the entire hierarchy of (antiplane viscoelasticity) bounds%
\index{bounds!hierarchy}
for two-dimensional transversely isotropic composites derived by \citeAPY{Milton:1981:BTO} are sharp. 

For two-component media, \citeAPY{Kantor:1984:IRB} obtained a representation formula%
\index{representation formula}
for the analytic properties of the effective elasticity 
tensor along a one-parameter trajectory%
\index{trajectory method}
in the moduli space (later generalized to two-parameter trajectories by \citeAPY{Ou:2012:TPI}).  A general framework for representation formulas, which yields representation formulas for the effective tensor for dielectrics, elasticity, piezoelectricity, thermoelasticity, thermoelectricity, and other coupled problems%
\index{coupled problems}
in multicomponent (possibly polycrystalline) non-lossy or lossy media (with possibly non-symmetric local tensors or having real and imaginary parts which do not necessarily commute)  was developed by Milton [see Chapters 18.6, 18.7 and 18.8 of \citeAY{Milton:2002:TOC}]. When more than two (real or complex) moduli are involved, another powerful approach, the field equation recursion method%
\index{field equation recursion method}
which is based on subspace collections, generates a whole hierarchy of bounds%
\index{bounds!hierarchy}
on effective tensors (not just on their associated quadratic forms), including the effective dielectric tensors of multicomponent (possibly polycrystalline) dielectric media with real or complex moduli, and the effective elastic or viscoelastic tensors of multicomponent (possibly polycrystalline) 
phases (\citeAY{Milton:1987:MCEa}, \citeyearNP{Milton:1987:MCEb}) [see also Chapter 29 of \citeAPY{Milton:2002:TOC}]. These bounds are applicable provided the 
real and imaginary parts of the local dielectric tensor, or viscoelasticity tensor, commute (i.e. can be simultaneously diagonalized in an appropriate basis).

Another breakthrough came when \citeAPY{Cherkaev:1994:VPC}%
\index{variational principles!Cherkaev and Gibiansky}
derived variational principles for electromagnetism with lossy materials and for viscoelasticity, assuming quasistatic equations and (fixed frequency) time harmonic fields. This provided a powerful tool for obtaining bounds on the complex dielectric constant of multicomponent (possibly anisotropic)  media \citePAY{Milton:1990:CSP} and for obtaining bounds on
the complex bulk and shear moduli of two- and three-dimensional two-phase composites (\citeAY{Gibiansky:1993:EVM}, Gibiansky and Lakes \citeyearNP{Gibiansky:1993:BCB}, \citeyearNP{Gibiansky:1997:BCB}, \citeAY{Milton:1997:EVM}, and \citeAY{Gibiansky:1999:EVM}), using both Hashin-Shtrikman method and the translation method. These variational principles of Cherkaev and Gibiansky have been extended to media
with non-symmetric tensors by \citeAPY{Milton:1990:CSP} [such as occur in conduction when a magnetic field is present: see \citeAY{Briane:2007:BSF}, where bounds are developed using these variational principles] and to beyond the quasistatic regime, to the full time harmonic equations of electromagnetism, acoustics, and elastodynamics in lossy inhomogeneous bodies 
(\citeAY{Milton:2009:MVP}; \citeAY{Milton:2010:MVP}).%
\index{variational principles!lossy media}

By contrast, very few results have been obtained regarding bounds on the creep and relaxation functions%
\index{bounds!creep functions}%
\index{bounds!relaxation functions}
in the time domain: \citeAPY{Schapery:1974:VBA} 
provided some interesting results via pseudo-elastic approximations;%
\index{pseudo-elastic approximations}
\citeAPY{Huet:1995:BOP} using the concept of a pseudo-convolutive bilinear form derived useful unilateral and bilateral bounds for the relaxation function tensor; \citeAPY{Vinogradov:2005:TCV}
obtained bounds which correlate the very short time response with the long time asymptotic behavior; and \citeAPY{Carini:2015:VFL} have derived some elementary bounds from their novel variational 
principles in the time domain,%
\index{variational principles!time domain}
which exploit the positive definiteness of a part of the constitutive law operator, combined with the transformation technique of 
\citeAPY{Cherkaev:1994:VPC} and \citeAPY{Milton:1990:CSP}  (note that Milton's work was based on that of Cherkaev and Gibiansky). The bounds of Carini and Mattei correlate the response at different times, while we are primarily interested in bounds on the response at a fixed given time. 
 
Here we use the analytic representation formula%
\index{representation formula}
developed by \citeAPY{Bergman:1978:DCC} (justified by \citeAPY{Milton:1981:BCP} and proved by \citeAPY{Golden:1983:BEP}) to obtain bounds on the macroscopic response of a two component composite (with microstructure independent of $x_1$) in the time domain for antiplane viscoelasticity. The key point which leads to the bounds is the observation that the response at a fixed time is linear in the residues (or eigenvalues of the residues when they are matrix valued) which enter the representation formula. This enables one to use linear programming theory to reduce the problem to one involving relatively few non-zero residues and then the optimization over the pole positions (and orientation of the residue matrices if they are anisotropic) can be done numerically.

There are two main conclusions that follow from our work. The first is that if sufficient information about the composite is incorporated in the bounds, such as the volume fractions of the phases and the
fact that the geometry is transversely isotropic, the bounds can be quite tight over the entire range of time. This should be very useful for predicting the transient behavior of composites. The second very 
significant point is that the bounds incorporating the volume fractions (and possibly transverse isotropy) can be extremely tight at certain specific times: thus measuring the response at such times, and using the bounds in an inverse fashion, could give very tight (and presumably useful) bounds on the volume fraction%
\index{bounds!volume fraction}
of the phases in the composite. The bounds we derive could be tightened further, for example, by incorporating information about the complex effective tensor measured at one or more frequencies (with cyclical loading). 

We remark that the method we use here is immediately applicable to bounding the transient response of three-dimensional two-component composites of lossy dielectric materials (or mixtures of a lossy material with a non-lossy one)(the case of two-dimensional two-component composites is of course mathematically isomorphic to the antiplane viscoelastic case studied here). This will be presented in a separate paper, directed towards physicists and electrical engineers. We also believe the method can be extended to obtain bounds on the transient response of fully  three-dimensional viscoelastic composites, not just in the antiplane case. In this setting, it is likely that the representation formulas for the effective elasticity tensor derived by \citeAPY{Kantor:1984:IRB} and \citeAPY{Ou:2012:TPI} and in Chapters 18.6, 18.7 and 18.8 of \citeAPY{Milton:2002:TOC}, or their generalizations, will prove useful.

\section{Summary of the results}\label{HResults}
\setcounter{equation}{0}

The results here presented concern bounds on the response, in terms of stresses and strains, of a two-component viscoelastic composite material in the time domain. We suppose that the external loadings are applied in such a way as to generate an \textit{antiplane shear}%
\index{antiplane shear}
state within the material. We recall that such a state is achieved when the components $u_2(\mathbf{x},t)$ and $u_3(\mathbf{x},t)$ of the displacement field $\mathbf{u}(\mathbf{x},t)$ are zero ($\mathbf{x}$ is the coordinate with respect to a Cartesian orthogonal reference system), 
for every $\mathbf{x}\in\Omega$ and every $t\in[0,+\infty)$, and the corresponding strain and stress states are of pure shear in the $12$- and $13$-planes, that is, by means of Voigt notation,%
\index{Voigt notation}
they are represented by the two-component vectors $\boldsymbol{\epsilon}(\mathbf{x},t)=\left[2{\epsilon}_{12}(\mathbf{x},t)\,\,\,2{\epsilon}_{13}(\mathbf{x},t)\right]^{\mathrm{T}}$ and $\boldsymbol{\sigma}(\mathbf{x},t)=\left[{\sigma}_{12}(\mathbf{x},t)\,\,\,{\sigma}_{13}(\mathbf{x},t)\right]^{\mathrm{T}}$. To ensure that a state of antiplane shear exists we assume that the microgeometry and, hence, the moduli depend only on $x_2$ and $x_3$.

We assume that both phases have an isotropic behavior, so that the direct and inverse constitutive laws, ruled by the $2\times 2$ matrices $\mathbf{C}(\mathbf{x},t)$ and $\mathbf{M}(\mathbf{x},t)$, read as follows
\begin{equation}\label{HConstitutive_law}
\boldsymbol{\sigma}(\mathbf{x},t)
=
\mathbf{C}(\mathbf{x},t)\ast
\boldsymbol{\epsilon}(\mathbf{x},t)
\quad\mbox{with} \quad \mathbf{C}(\mathbf{x},t)=\sum_{i=1,2}\chi_i(\mathbf{x})\,\mu_i(t)\mathbf{I},
\end{equation}
\begin{equation}\label{HInverse_Constitutive_law}
\boldsymbol{\epsilon}(\mathbf{x},t)
=
\mathbf{M}(\mathbf{x},t)\ast
\boldsymbol{\sigma}(\mathbf{x},t)
\quad\mbox{with} \quad \mathbf{M}(\mathbf{x},t)=\sum_{i=1,2}\chi_i(\mathbf{x})\,\zeta_i(t)\mathbf{I},
\end{equation}
where $\ast$ indicates a time convolution,%
\index{convolution}
$\mathbf{I}$ is the identity matrix, $\chi_i(\mathbf{x})$ is the indicator function of phase $i$, and $\mu_i(t)$ and $\zeta_i(t)$ are, respectively, the  shear stiffness and the shear compliance of phase $i$, both functions of time. A word about the notation may be helpful: on the left hand side of \eqref{HConstitutive_law} [and
\eqref{HInverse_Constitutive_law}] $\boldsymbol{\sigma}(\mathbf{x},t)$ [respectively $\boldsymbol{\epsilon}(\mathbf{x},t)$] refers to the stress [strain] at a specific time $t$, while
on the right hand side $\mathbf{C}(\mathbf{x},t)$ and $\boldsymbol{\epsilon}(\mathbf{x},t)$ [$\mathbf{M}(\mathbf{x},t)$ and $\boldsymbol{\sigma}(\mathbf{x},t)$] 
refer to the relaxation kernel%
\index{relaxaton kernel}
and strain [creep kernel%
\index{creep kernel}
and stress] as functions of time from time 0 (before which there is no stress or strain) up to time $t$, which are convolved
together to produce the stress [strain] at the specific time $t$. 

In this investigation, we are interested in determining the effective behavior of the composite (we consider the most general case for which the composite does not have any specific symmetry), described by the effective direct and inverse constitutive law operators ${\mathbf{C}_*(t)}$ and $\mathbf{M}_*(t)$ as follows
\begin{equation}\label{HEffective_Constitutive_Laws}
{\overline{\boldsymbol{\sigma}}(t)}={\mathbf{C}_*(t)}\ast{\overline{\boldsymbol{\epsilon}}(t)},\quad\quad\quad
{\overline{\boldsymbol{\epsilon}}(t)}={\mathbf{M}_*(t)}\ast{\overline{\boldsymbol{\sigma}}(t)},
\end{equation}
where here and henceforth the bar denotes the volume average operation. In particular, we seek estimates for the shear stress and strain components $\overline{\sigma}_{12}(t)$ and $\overline{\epsilon}_{12}(t)$ for each time $t\in[0,\infty)$. 

By applying the so-called \textit{analytic method},%
\index{analytic method}
based on the analyticity properties of the Laplace transforms ${\mathbf{C}_*(\lambda)}$ and $\mathbf{M}_*(\lambda)$ ($\lambda$ is the Laplace transform parameter) of the operators ${\mathbf{C}_*(t)}$ and $\mathbf{M}_*(t)$ as functions of the Laplace transforms $\mu_i(\lambda)$ and $\zeta_i(\lambda)$ of $\mu_i(t)$ and $\zeta_i(t)$, $i=1,2$ (see Section \ref{HProblem}), the effective constitutive laws \eqref{HEffective_Constitutive_Laws} turn into:
\begin{equation}\label{Hsigma_con_trasformata}
\overline{\boldsymbol{\sigma}}(t)=\mu_2(t)\ast\overline{\boldsymbol{\epsilon}}(t)- \sum_{i=0}^m\mathbf{B}_{i}\,\mathcal{L}^{-1}\left[\frac{\mu_2(\lambda)}{s(\lambda)-s_{i}}\right](t)\ast\overline{\boldsymbol{\epsilon}}(t),
\end{equation}
\begin{equation}\label{Hepsilon_con_trasformata}
\overline{\boldsymbol{\epsilon}}(t)=\zeta_2(t)\ast\overline{\boldsymbol{\sigma}}(t)- \sum_{i=0}^m\mathbf{P}_{i}\,\mathcal{L}^{-1}\left[\frac{\zeta_2(\lambda)}{u(\lambda)-u_{i}}\right](t)\ast\overline{\boldsymbol{\sigma}}(t),
\end{equation}
where $\mathcal{L}^{-1}$ represents the inverse of the Laplace transform, $s_{i}$ and $u_i$ are, respectively, the poles of the functions 
\begin{equation}\label{HF_G_definition}
\mathbf{F}(s)=\mathbf{I}-\frac{\mathbf{C}_*(\lambda)}{{\mu}_2(\lambda)},\quad\quad\quad\mathbf{G}(u)=\mathbf{I}-\frac{\mathbf{M}_*(\lambda)}{{\zeta}_2(\lambda)},
\end{equation}
with residues $\mathbf{B}_{i}$ and $\mathbf{P}_{i}$, respectively, where the parameters $s(\lambda)$ and $u(\lambda)$ are defined as follows
\begin{equation}\label{Hs_u}
s(\lambda)=\frac{{\mu}_2(\lambda)}{{\mu}_2(\lambda)-{\mu}_1(\lambda)},\quad\quad\quad
u(\lambda)=\frac{{\zeta}_2(\lambda)}{{\zeta}_2(\lambda)-{\zeta}_1(\lambda)}.
\end{equation}
The poles $s_i$ and $u_i$ lie on the semi-closed interval $[0,1)$ and the residues $\mathbf{B}_i$ and $\mathbf{P}_{i}$ are positive semi-definite matrices. It must be noted that equations \eqref{Hsigma_con_trasformata} and \eqref{Hepsilon_con_trasformata} hold only in case ${\mathbf{C}_*(\lambda)}$ and $\mathbf{M}_*(\lambda)$ are rational functions of the eigenvalues $\mu_i(\lambda)$ and $\zeta_i(\lambda)$, $i=1,2$, respectively. There is no lack of generality in considering only rational functions,%
\index{approximation!rational functions}
since irrational functions can be approximated to an arbitrarily high
degree of approximation by rational ones, except in the near vicinity of their poles.

All the information about the composite, such as the knowledge of the volume fractions or the eventual isotropy of the material, is then transformed into constraints on the residues $\mathbf{B}_{i}$ and $\mathbf{P}_{i}$, the so-called \textit{sum rules},%
\index{sum rules}
introduced by \citeAPY{Bergman:1978:DCC} and  discussed in Section \ref{HProblem}. Such constraints are then contextualized in Section \ref{HSum rules} so that bounds on the components $\overline{\sigma}_{12}(t)$ and $\overline{\epsilon}_{12}(t)$ are derived by means of the theory of linear programming. In particular, for each information available about the composite, that is, for each sum rule that is taken into account, we provide analytic expressions for the maximum and minimum values of the field components $\overline{\sigma}_{12}(t)$ and $\overline{\epsilon}_{12}(t)$
at each instant in time, when the applied fields are respectively $\overline{\boldsymbol{\epsilon}}(t)=[\overline{\epsilon}_{12}(t)\,\,\, 0]^T$ and 
$\overline{\boldsymbol{\sigma}}(t)=[\overline{\sigma}_{12}(t)\,\,\,  0]^T$ (see Section \ref{Hbounds}). 

In this section we present some numerical results by specifying the models used for the behavior of the two phases, so that the inverse of the Laplace transform in \eqref{Hsigma_con_trasformata} and \eqref{Hepsilon_con_trasformata} can be calculated explicitly.

\subsection{Bounds on the stress response}%
\index{bounds!stress response}
 
For the sake of simplicity, we suppose that phase 2 is characterized by a linear elastic behavior, with shear modulus $\mu_2(t)=G_2\delta(t)$, $\delta(t)$ being the Dirac delta function, and that phase 1 is described by the Maxwell model.%
\index{Maxwell model}
We recall that such a model is represented by a purely Newtonian viscous damper%
\index{viscous damper}
(viscosity coefficient $\eta_M$) and a purely Hookean elastic spring (elastic modulus $G_M$) connected in series so that the shear modulus of phase 1 is $\mu_1(t)=G_M\delta(t)-G_M^2/\eta_M\mathrm{exp}[-G_M t/\eta_M]$. To capture the most interesting case, we suppose that 
the material is not ``well-ordered",%
\index{non-well-ordering}
that is, the product of the difference of the instantaneous moduli (very close to $t=0$) and the long time moduli (as $t$ tends to infinity) is negative, i.e., $G_2<G_M$. Nevertheless, for completeness, in the following we will show also some results concerning the ``well-ordered"%
\index{well-ordering}
case, that is, when the product of these differences is positive ($G_2>G_M$).

We consider the classical relaxation test in which the applied average stain is held constant after being initially 
applied, i.e. $\overline{\boldsymbol{\epsilon}}(t)= \boldsymbol{\epsilon}_0=[\epsilon_0\,\,\,0]^{\mathrm{T}}$. From \eqref{Hsigma_con_trasformata} we derive the following expression for $\overline{\sigma}_{12}(t)$:
\begin{equation}\label{Hsigma12_t}
\overline{\sigma}_{12}(t)=G_2{\epsilon}_0-G_2{\epsilon}_0\sum_{i=0}^m\left\{1-\frac{\mathrm{exp}\left[-\frac{G_2(1-s_i)t}{\eta_M\left(\frac{G_2}{G_M}-s_i\left(\frac{G_2}{G_M}-1\right)\right)}\right]}{\frac{G_2}{G_M}-s_i\left(\frac{G_2}{G_M}-1\right)}\right\}\frac{{B}_{11}^{(i)}}{1-s_i},
\end{equation}
where ${B}_{11}^{(i)}$ are the $11$-components of the $2\times 2$ matrices $\mathbf{B}_i$.

Now suppose that no information about the geometry of the composite is available. As shown in Subsection \ref{HBounds_stress}, in order to optimize $\overline{\sigma}_{12}(t)$ for any given time $t\in[0,\infty)$, it suffices (by linear programming theory)%
\index{linear programming theory}
 to take only one element ${B}_{11}^{(0)}$ to be non zero. In particular, it turns out that ${B}_{11}^{(0)}=1-s_0$ and the expression of $\overline{\sigma}_{12}(t)$ is then given by 
\begin{equation}\label{Hlamresp}
\overline{\sigma}_{12}(t)=G_2\epsilon_0\frac{\mathrm{exp}\left[-\frac{G_2(1-s_0)t}{\eta_M [s_0\left(1-G_2/G_M\right)+G_2/G_M]}\right]}{s_0\left(1-G_2/G_M\right)+G_2/G_M}.
\end{equation} 
The maximum (or minimum) value of $\overline{\sigma}_{12}(t)$ is obtained by varying the pole $s_0$ over its domain of validity, i.e. $[0,1)$. Since the response \eqref{Hlamresp} corresponds to that
of a laminate oriented with the $x_2$ axis normal to the layer planes, varying $s_0$ corresponds to varying the volume fraction of the phases in the laminate%
\index{laminate}
(since no information about the composite is available, the volume fraction $f_1$ of phase 1 can be varied from $0$ to $1$).
In particular, the case $s_0=0$ corresponds to a ``composite'' which contains only  phase 1, while the case $s_0\to 1$ corresponds to a ``composite'' which contains only  phase 2. 

\begin{figure}[htbp]
\centering
\includegraphics[width=0.6\textwidth]{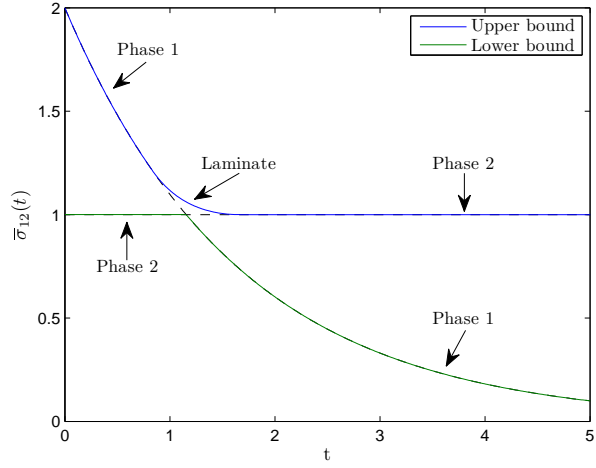}
\caption[Lower and upper bounds on $\overline{\sigma}_{12}(t)$ in case no information about the composite is given.]{Lower and upper bounds on $\overline{\sigma}_{12}(t)$ in case no information about the composite is given. The stress $\overline{\sigma}_{12}(t)$ is normalized with respect to the elastic stress in phase 2, equal to $\epsilon_0G_2$. The material purely made of phase 1 provides the upper bound for $t\leq t_1=0.83$ and the lower bound for $t\geq t_2=1.15$, whereas the material purely made of phase 2 attains the lower bound for $t\leq t_2=1.15$ and the upper for $t\geq t_1=1.67$. For $t_1\leq t\leq t_3$ the upper bound is realized by a laminate of the two components.}
\label{Hfig:Max_noinfo}
\end{figure}

As shown in Fig. \ref{Hfig:Max_noinfo}, where $\overline{\sigma}_{12}(t)$ is normalized with respect to the stress state in the elastic phase, $G_2\epsilon_0$, the material purely made of phase 1 ($s_0=0$) attains the upper bound for $t\leq t_1=\eta_M/G_M(1-G_2/G_M)$ (equal to $0.83$ in Fig. \ref{Hfig:Max_noinfo}) and the lower bound for $t\geq t_2=\eta_M/G_M\log(G_M/G_2)$ (equal to $1.15$ in Fig. \ref{Hfig:Max_noinfo}), whereas the material purely made of phase 2 ($s_0\to 1$) attains the lower bound for $t\leq t_2$ and the upper one for $t\geq t_3=\eta_M/G_2(1-G_2/G_M)$ (equal to $1.67$ in Fig. \ref{Hfig:Max_noinfo}): the same microstructure can provide both the maximum and the minimum response depending on the interval of time considered. Furthermore, for $t_1\leq t\leq t_3$ the upper bound is realized by a laminate of the two components corresponding to the pole $s_0$ positioned at 
\begin{equation}\label{Hs_0opt}
s_0^{opt}=\frac{\frac{tG_2}{\eta_K}-\frac{G_2}{G_K}\left(1-\frac{G_2}{G_K}\right)}{\left(1-\frac{G_2}{G_K}\right)^2}.
\end{equation} 
Due to the dependence of $s_0^{opt}$ on time $t$, it follows that the volume of the phases in the laminate attaining the bounds 
needs to be adjusted according to the time at which one is optimizing the response.

Specifically, the upper bound is given by
\begin{equation}\label{Hsigma_max}
\overline{\sigma}_{12}^{\max}(t)=
\left\{
\begin{array}{lll}
\epsilon_0G_M\mathrm{exp}\left[-\frac{G_M}{\eta_M}t\right]&s_0=0&,t\leq t_1,\\
\epsilon_0\frac{\eta_M}{t}\left(1-\frac{G_2}{G_M}\right)\mathrm{exp}\left[1-\frac{t}{\eta_M}\frac{G_2G_M}{G_M-G_2}\right] &s_0=s_0^{opt}&,t_1\leq t \leq t_3,\\
\epsilon_0G_2&s_0\to 1&, t \geq t_3,
\end{array}
\right.
\end{equation} 
whereas the lower bound corresponds to
\begin{equation}\label{Hsigma_min}
\overline{\sigma}_{12}^{\min}(t)=\left\{
\begin{array}{lll}
\epsilon_0G_2&s_0\to 1&,t \leq t_2,\\
\epsilon_0G_M\mathrm{exp}\left[-\frac{G_M}{\eta_M}t\right]&s_0=0&,t\geq t_2.
\end{array}
\right.
\end{equation} 

\begin{figure}[htbp]
\centering
\includegraphics[width=0.6\textwidth]{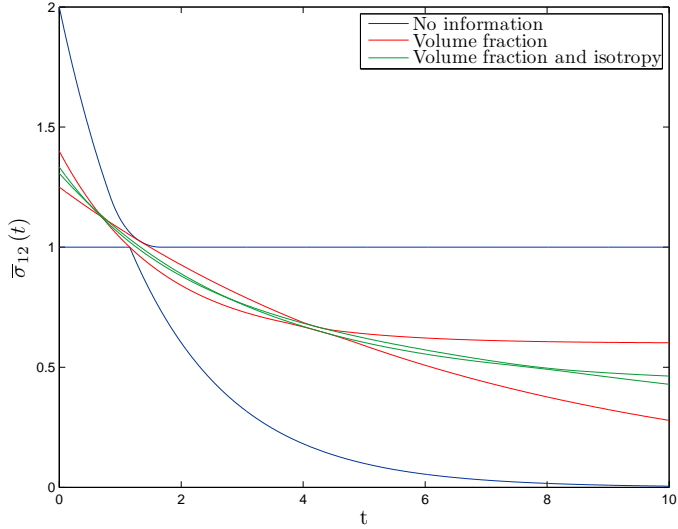}
\caption[The lower and upper bounds on $\overline{\sigma}_{12}(t)$ with no information, volume fraction information, and with isotropy plus known volume fractions.]{Comparison between the lower and upper bounds on $\overline{\sigma}_{12}(t)$ (normalized with respect to the elastic stress in phase 2, equal to $\epsilon_0G_2$) in the following three cases: no information about the composite is given; the volume fraction of the components is known ($f_1=0.4$); and the composite is isotropic with given volume fractions. The bounds become tighter and tighter as more information on the composite structure is included.}
\label{Hfig:Max_r0f1isotropy}
\end{figure}

In case the volume fractions of the components are known, tighter bounds can be obtained. In particular, in Fig. \ref{Hfig:Max_r0f1isotropy} we compare the results obtained by considering the following situations: no information about the composite is available (the case analyzed in detail above), the volume fraction of the constituents is known (two poles), and the composite is transversely isotropic with given volume fractions (three poles). It is worth noting that the bounds corresponding to the latter case are very tight and therefore the response of the composite in terms of $\overline{\sigma}_{12}(t)$ is almost completely determined. 

Significantly, the bounds in Fig. \ref{Hfig:Max_r0f1isotropy} which include the volume fraction (and possibly, transverse isotropy) are extremely tight at particular times $t$, and so, if the volume fraction is unknown, we can measure the value of $\overline{\sigma}_{12}(t)$ at these times, and then use the bounds in an inverse fashion to determine (almost exactly) the volume fraction.%
\index{bounds!volume fraction}
To understand why the bounds are extremely tight at these times we rewrite the relation \eqref{Hsigma12_t} in the form

\begin{equation}\label{Hsep}
\overline{\sigma}_{12}(t)=G_2{\epsilon}_0-G_2{\epsilon}_0\sum_{i=0}^mK(s_i,t){B}_{11}^{(i)},
\end{equation}
with coefficients
\begin{equation}
K(s_i,t)=\left\{1-\frac{\mathrm{exp}\left[-\frac{G_2(1-s_i)t}{\eta_M\left(\frac{G_2}{G_M}-s_i\left(\frac{G_2}{G_M}-1\right)\right)}\right]}
{\frac{G_2}{G_M}-s_i\left(\frac{G_2}{G_M}-1\right)}\right\}\frac{1}{1-s_i}.
\end{equation}
If at a time $t=\tau_0$ the coefficients $K(s_i,t)$ were almost independent of $s_i$, i.e., $K(s_i,\tau_0)\approx c_0$ for all $i$, then by substituting this in \eqref{Hsep} and using the sum rule 
given later in \eqref{Hconstraint1_bA}, we see that 
\begin{equation}
\overline{\sigma}_{12}(\tau_0)\approx G_2{\epsilon}_0-G_2{\epsilon}_0c_0f_1.
\end{equation}
Alternatively, if at another time $t=\tau_1$ the coefficients $K(s_i,t)$ depend almost linearly on $s_i$, i.e., $K(s_i,\tau_1)\approx c_0+c_1s_i$ for all $i$, and the geometry is transversely isotropic,
then by substituting this in \eqref{Hsep} and using the sum rules 
given later in \eqref{Hconstraint1_bA} and  \eqref{Hconstraint2_bA} we see that 
\begin{equation}
\overline{\sigma}_{12}(\tau_1)\approx G_2{\epsilon}_0-G_2{\epsilon}_0(c_0f_1+c_1f_1f_2/2).
\end{equation}
Video 1 shows $K(s_i,t)$ as a function of time for our example, and we see indeed that the coefficients $K(s_i,t)$ are almost independent of $s_i$ at the times when the bounds which incorporate only
the volume fraction are very tight (for example, at $\tau_0=0.78$ and at $\tau_0=4.3$ - see also Fig. \ref{Hfig:Max_r0f1isotropy}), and they depend almost linearly on $s_i$ at the times when the bounds which incorporate the volume fractions and the transverse isotropy are very tight (for instance, at $\tau_1=2.8$ and $\tau_1=8.21$ - see also Fig. \ref{Hfig:Max_r0f1isotropy}).

\begin{figure}[htbp]
\centering
\includegraphics[width=0.6\textwidth]{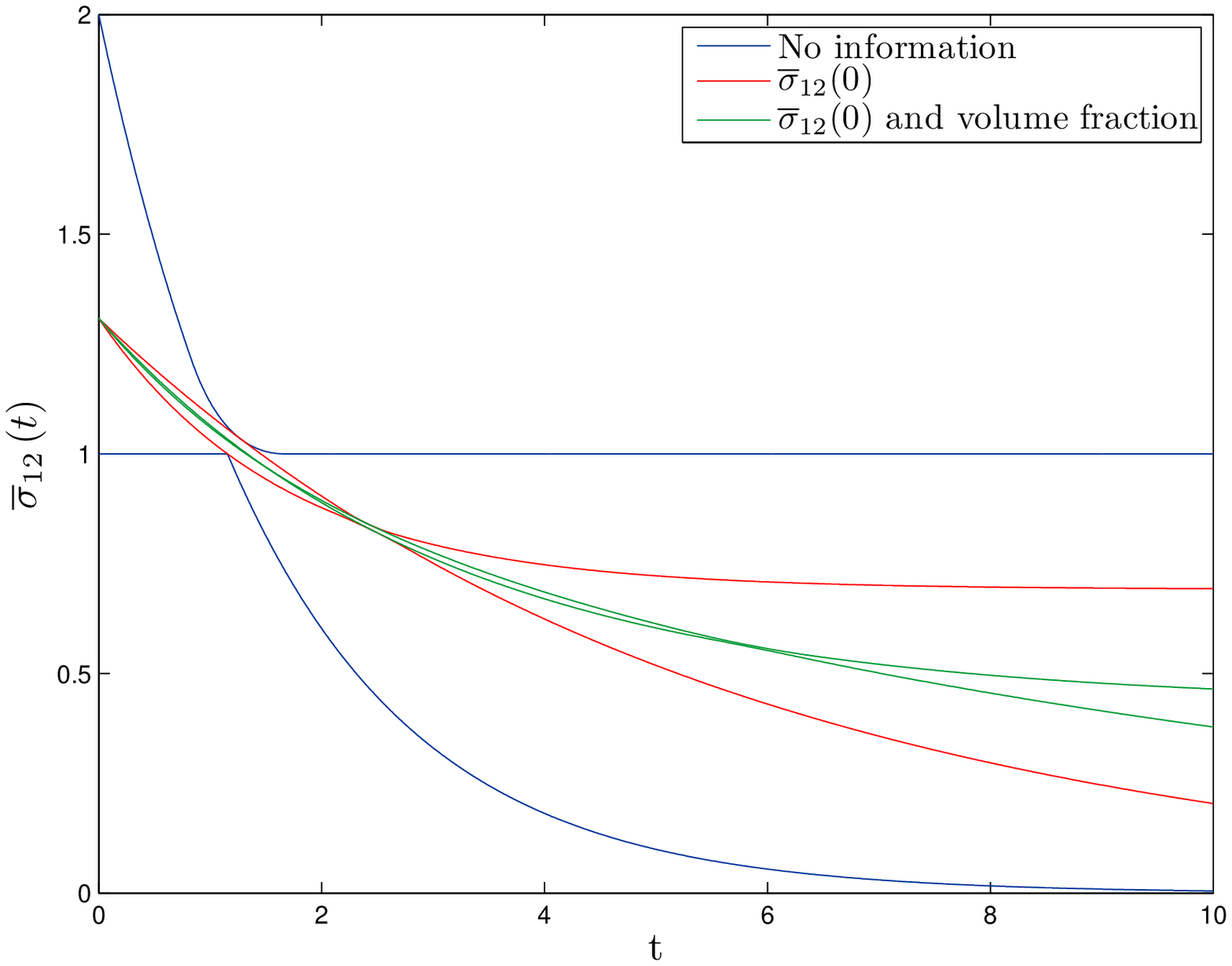}
\caption[The lower and upper bounds on $\overline{\sigma}_{12}(t)$ with no information, known $\overline{\sigma}_{12}(0)$, and known $\overline{\sigma}_{12}(0)$ plus known volume fractions.]
        {Comparison between the lower and upper bounds on $\overline{\sigma}_{12}(t)$ (normalized with respect to the elastic stress in phase 2, equal to $\epsilon_0G_2$) in the following three cases: no information about the composite is given; the value of $\overline{\sigma}_{12}(t)$ at $t=0$ is prescribed; and the value of $\overline{\sigma}_{12}(t)$ at $t=0$ and the volume fractions are known ($f_1=0.4$). The bounds become tighter and tighter as more information about the composite structure is included.}
\label{Hfig:Max_r0f1sigma0}
\end{figure}

\begin{figure}[htbp]
\centering
\includegraphics[width=0.6\textwidth]{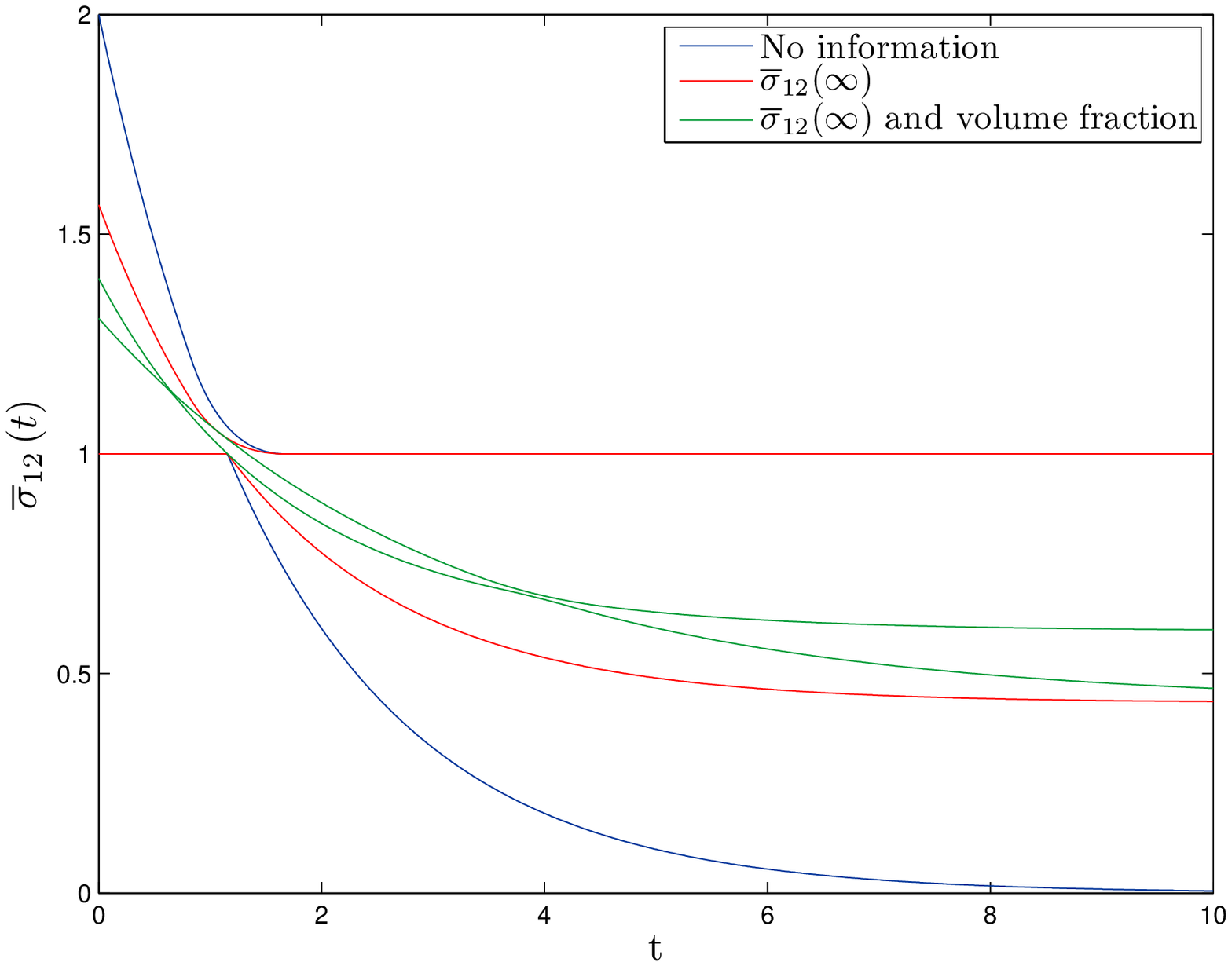}
\caption[The lower and upper bounds on $\overline{\sigma}_{12}(t)$ with no information, known $\overline{\sigma}_{12}(\infty)$, and known $\overline{\sigma}_{12}(\infty)$ plus known volume fractions.]{Comparison between the lower and upper bounds on $\overline{\sigma}_{12}(t)$ (normalized with respect to the elastic stress in phase 2, equal to $\epsilon_0G_2$) in the following three cases: no information about the composite is given; the value of $\overline{\sigma}_{12}(t)$ at $t\to\infty$ is prescribed; and the value of $\overline{\sigma}_{12}(t)$ at $t\to\infty$ and the volume fractions are known ($f_1=0.4$). In the last two cases, the upper bound attains the assigned value of $\overline{\sigma}_{12}(t)$ at $t\to\infty$ only in the near vicinity of $t\to\infty$, whereas the lower bound converges very fast.}
\label{Hfig:Max_r0f1sigmainf}
\end{figure}

Other information about the composite can be considered, such as the knowledge of the value of $\overline{\sigma}_{12}(t)$ at a specific time. Figs. \ref{Hfig:Max_r0f1sigma0} and \ref{Hfig:Max_r0f1sigmainf} show the results obtained in case the value of $\overline{\sigma}_{12}(t)$ at $t=0$ and $t\to\infty$, respectively, is given. 


With reference to Fig. \ref{Hfig:Max_r0f1sigma0}, notice that the combination of the knowledge of the volume fraction and of the value of $\overline{\sigma}_{12}(t)$ at $t=0$, $\overline{\sigma}_{12}(0)$, provides very tight bounds on $\overline{\sigma}_{12}(t)$.

Concerning Fig. \ref{Hfig:Max_r0f1sigmainf}, a few remarks should be made. First of all, notice that for the case when only the value of $\overline{\sigma}_{12}(t)$ for $t\to\infty$, $\overline{\sigma}_{12}(\infty)$, is prescribed, the upper bound seems to not reach such a value: it provides a constant stress state equal to the one in the material purely composed of phase 2. This is due to the fact that the only non-zero residue $B_{{11}}^{(0)}=(1-s_0)\left(1-\overline{\sigma}_{12}(\infty)/(G_2\epsilon_0)\right)$ in \eqref{Hsigma12_t} takes a value very close to zero when the corresponding pole $s_0$ tends to $1$ and, consequently, the predicted response is that of phase 2 (see equation \eqref{Hsigma12_t}). However, in the near vicinity of $t\to \infty$, the upper bound rapidly converges to the prescribed value $\overline{\sigma}_{12}(\infty)$. Regarding the upper bound obtained by considering both the values of $\overline{\sigma}_{12}(t)$ at $t\to\infty$ and the volume fractions to be known, notice that it presents a small slope which allows it to slowly reach the value $\overline{\sigma}_{12}(\infty)$ for $t\to\infty$.

\begin{figure}[htbp]
\centering
\includegraphics[width=0.6\textwidth]{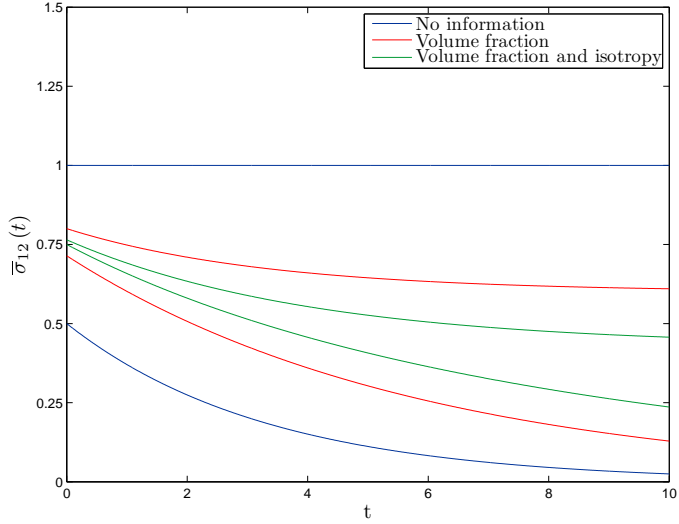}
\caption[The lower and upper bounds on $\overline{\sigma}_{12}(t)$ in the ``well-ordered case'' $G_2>G_M$ with no information, volume fraction information, and with isotropy plus known volume fractions.]{Comparison between the lower and upper bounds on $\overline{\sigma}_{12}(t)$ (normalized with respect to the elastic stress in phase 2, equal to $\epsilon_0G_2$) in the ``well-ordered case'' $G_2>G_M$. The following three subcases are considered: no information about the composite is given; the volume fraction of the components is known ($f_1=0.4$); and the composite is isotropic with given volume fractions. The bounds become tighter as more information on the composite structure is included, but remain quite wide except near $t=0$.}
\label{Hfig:Max_r0f1isotropy_mod}
\end{figure}

The ``well ordered" case,%
\index{well-ordering}
corresponding to the choice $G_2>G_M$, is less interesting due to the fact that the curves representing the behavior of phase 1 and phase 2 do not intersect. However, for completeness, in Fig. \ref{Hfig:Max_r0f1isotropy_mod}, we provide bounds on $\overline{\sigma}_{12}(t)$ for $G_2>G_M$ in the following cases: no information about the composite is available; the volume fraction is known; and the composite is transversely isotropic with given volume fraction. Again, the bounds become tighter the more information about the composite is considered. Nevertheless, the bounds are wide compared to the case $G_2<G_M$, and are tightest at $t=0$.\\
\\
Besides optimizing the component $\overline{\sigma}_{12}(t)$ of the averaged stress field $\overline{\boldsymbol{\sigma}}(t)$, one would like to determine also what
are the possible values the vector $\overline{\boldsymbol{\sigma}}(t)=[\overline{\sigma}_{12}(t)\,\,\, \overline{\sigma}_{13}(t)]^T$%
\index{bounds!stress vector}
can take as time evolves. One way to get some information about this
is to look for the maximum or minimum value attained by a linear combination of the components $\overline{\sigma}_{12}(t)$ and $\overline{\sigma}_{13}(t)$ of $\overline{\boldsymbol{\sigma}}(t)$. Let us consider, then, the following scalar objective function,%
\index{objective function}
for each fixed angle $\alpha$:  
\begin{equation}\label{HF}
\mathcal{F}(t)=\sin\alpha\,\overline{\sigma}_{12}(t)+\cos\alpha\,\overline{\sigma}_{13}(t),
\end{equation}
where, in general, $\overline{\sigma}_{12}(t)$ and $\overline{\sigma}_{13}(t)$ are given by \eqref{Hsigma_con_trasformata}. 

Let us assume that the same hypotheses valid for the bounds on $\overline{\sigma}_{12}(t)$ still hold, i.e., phase 1 is described by the Maxwell model, phase 2 is elastic, and $\overline{\boldsymbol{\epsilon}}(t)=\boldsymbol{\epsilon}_0= [\epsilon_0\,\,\,0]^{\mathrm{T}}$. Then, we have
\begin{equation}\label{Hsigma_t}
\overline{\boldsymbol{\sigma}}(t)=G_2\boldsymbol{\epsilon}_0-G_2\sum_{i=0}^m\left\{1-\frac{\mathrm{exp}\left[-\frac{G_2(1-s_i)t}{\eta_M\left(\frac{G_2}{G_M}-s_i\left(\frac{G_2}{G_M}-1\right)\right)}\right]}{\frac{G_2}{G_M}-s_i\left(\frac{G_2}{G_M}-1\right)}\right\}\frac{\mathbf{B}_i}{1-s_i}\boldsymbol{\epsilon}_0.
\end{equation}
Furthermore, we suppose that the microstructure has \textit{reflective symmetry},%
\index{reflective symmetry}
that is, it is symmetric with respect to reflection about a certain plane. Such an assumption implies that all residues $\mathbf{B}_i$ in \eqref{Hsigma_t} are diagonal matrices with respect to the same basis (i.e., they commute). In general, optimizing the quantity $\mathcal{F}(t)$ for a fixed $\alpha$ and, then, varying $\alpha$ will only allow us
to find the convex hull of the set of possible vectors $\overline{\boldsymbol{\sigma}}(t)$ at each time $t$. However, in the case of reflective symmetry, we can first fix the orientation
of the residues (i.e., the orientation of the composite) \footnote{Fixing the orientation of the composite means fixing the value of the angle $\theta=\theta_i$, for each $i$, in equation \eqref{Htheta_i}. In particular, when the orientation is fixed, two possible configurations of the microstructure are admissible: one corresponds to the angle $\theta$ and the other, reflected with respect to the first one, corresponds to the angle $\theta+\pi/2$. Strictly speaking
the microstructure does not necessarily have this additional reflective symmetry, but the associated effective tensor does have it.} and, then, we can find the minimum value of $\mathcal{F}(t)$, and a (possibly non-unique) function $\overline{\boldsymbol{\sigma}}(t,\alpha)$
which realizes it (observe that finding the maximum value of $\mathcal{F}(t)$ is the same as finding the minimum when $\alpha$ is replaced by $\alpha+\pi$). Next, for each $t$ construct the 
set which is the union of the points $\overline{\boldsymbol{\sigma}}(t,\alpha)$ as $\alpha$ varies between $0$ and $2\pi$, and take its convex hull: the boundary of this convex hull 
is the trajectory of $\overline{\boldsymbol{\sigma}}(t,\alpha)$ as $\alpha$ increases, except if there is a jump in the value of $\overline{\boldsymbol{\sigma}}(t,\alpha)$, for which the 
successive values of $\overline{\boldsymbol{\sigma}}(t,\alpha)$ are joined by a straight line. Finally, we take the union of these convex hulls as the orientation is varied. In this way we obtain bounds which at any instant of time $t$ confine the pair $\left(\overline{\sigma}_{13}(t), \overline{\sigma}_{12}(t)\right)$ to a region which is not necessarily convex.

In case no information about the geometry of the
composite is available, apart from the reflective symmetry, the optimum value of $\mathcal{F}(t)$ is attained when a maximum of two residues are non-zero. 

Video 2, plots $\overline{\sigma}_{12}(t)$ against $\overline{\sigma}_{13}(t)$ (both normalized by $G_2\epsilon_0$, the stress state in phase 2) for each moment of time, in case the orientation of the composite is fixed (blue curve). To enrich the results, the video also plots the domain $\left(\overline{\sigma}_{13}(t),\overline{\sigma}_{12}(t)\right)$ corresponding to the stress state in a laminate%
\index{laminate}
with the prescribed orientation (red curve). We recall that for a laminate the stress state is unequivocally determined, since the eigenvalues of the two non-null residues are related to the harmonic and arithmetic means of the moduli of the two phases. Note that, since no information about the composite is available, the volume fraction $f_1$ of phase $1$ can vary from $0$ to $1$. 
%
In the initial frame of the video, at $t = 0$, the point $(0,1)$ (corresponding to $s_0\to 1,\,s_1\to1$) represents the instantaneous stress state within phase 2, whereas the point $(0,2)$ (corresponding to $s_0=s_1=0$) represents the stress state within phase 1. Obviously, both points belong also to the red curve representing the laminate behavior. As time goes by, the domain becomes smaller and smaller with the upper vertex still representing the behavior of phase 1 while the lower vertex, the point $(0,1)$, remaining fixed as it represents the elastic behavior of phase 2. For times $t$ between $t_1=0.83$ and $t_3=1.67$ a change takes place: the upper vertex 
does not represent the response of a phase 1, nor even that of a laminate. Then for times $t>t_3=1.67$, the upper vertex coincides 
with the point $(0,1)$, representing the behavior of phase 2. The lower vertex describes the behavior of phase 2 until $t=t_2=1.15$, after which it 
represents the behavior of phase 1.

When the orientation of the composite is not known, one has to perform the previous analysis for each possible orientation and, then, take the union of the resulting domains, as shown in Video 3.

For each fixed value of the angle $\alpha$, sharp bounds on the function $\mathcal{F}(t)$ \eqref{HF} give the
straight lines forming an angle equal to $\alpha$, with respect to the $\overline{\sigma}_{13}(t)$-axis, which are tangent
to the domain of possible $\left(\overline{\sigma}_{13}(t),\overline{\sigma}_{12}(t)\right)$. For each time, the values of 
$\left(\overline{\sigma}_{13}(t),\overline{\sigma}_{12}(t)\right)$ which attain the bounds on $\mathcal{F}(t)$ correspond
to those points where the tangent line intersects this domain.

We do not provide numerical results bounding the function $\mathcal{F}(t)$ for the case in which the volume fractions of
the components are known, due to the large number of variables involved.

\subsection{Bounds on the strain response}%
\index{bounds!strain response}

In this case, we suppose that phase 2 is still elastic, with $\zeta_2(t)=1/G_2\delta(t)$, while we represent the behavior of phase 1 by means of the Kelvin-Voigt model,%
\index{Kelvin-Voigt model}
composed by a purely viscous damper%
\index{viscous damper}
($\eta_K$) and purely elastic spring ($G_K$) connected in parallel, so that $\zeta_1(t)=\mathrm{exp}(-G_Kt/\eta_K)/\eta_K$. The most interesting results correspond to the non ``well-ordered" case,%
\index{non-well-ordering}
corresponding to $G_2<G_K$. 

Moreover, if we consider the classical creep test, for which the applied averaged stress field is constant in time after
it has been initially imposed, i.e., $\overline{\boldsymbol{\sigma}}(t)=\boldsymbol{\sigma}_0$, and we set $\boldsymbol{\sigma}_0=[\sigma_0\,\,\,0]^{\mathrm{T}}$, then equation \eqref{Hepsilon_con_trasformata} yields:
\begin{equation}\label{Hepsilon12_t}
\overline{\epsilon}_{12}(t)=\frac{\sigma_0}{2G_2}-\frac{\sigma_0}{2G_2}\sum_{i=0}^m\left\{G_K-G_2+G_2\frac{\mathrm{exp}\left[-\frac{(G_K(1-u_i)+u_iG_2)t}{\eta_K\left(1-u_i\right)}\right]}{1-u_i}\right\}\frac{{P}_{11}^{(i)}}{G_K-u_i(G_K-G_2)},
\end{equation}
where $P_{11}^{(i)}$ are the $11$-components of the residues $\mathbf{P}_i$. 

In case no information about the geometry of the composite is available, bounds on $\overline{\epsilon}_{12}(t)$ are obtained by taking only one residue to be non-zero (see Subsection \ref{HBounds_strain}). In particular, it turns out that $P_{11}^{(0)}=1-u_0$ and $\overline{\epsilon}_{12}(t)$, from \eqref{Hepsilon12_t}, takes the following expression:
\begin{equation}
\overline{\epsilon}_{12}(t)=\frac{\sigma_0}{2G_2}\left\{1-\frac{(1-u_0)(G_K-G_2)+G_2\mathrm{exp}\left[-\frac{G_K-u_0(G_K-G_2)}{\eta_K(1-u_0)}t\right]}{G_K-u_0(G_K-G_2)}\right\}.
\end{equation}
\begin{figure}[htbp]
\centering
\includegraphics[width=0.6\textwidth]{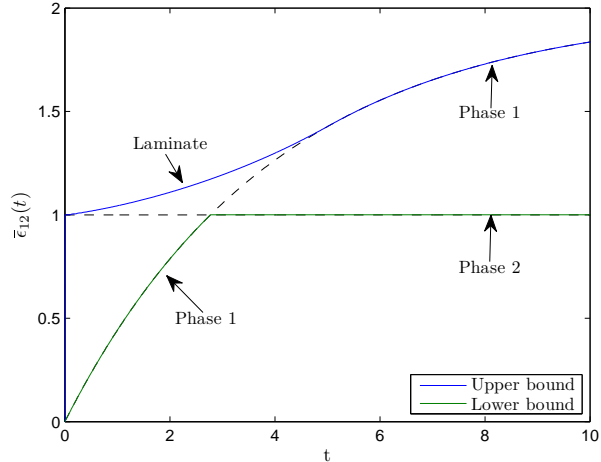}
\caption[Lower and upper bounds on $\overline{\epsilon}_{12}(t)$ in case no information about the composite is given.]{Lower and upper bounds on $\overline{\epsilon}_{12}(t)$ (normalized with respect to the elastic strain in phase 2, equal to $\sigma_0/(2G_2)$) in case no information about the composite is given. The material purely made of phase 1 provides the lower bound for $t\leq t_{I}=2.84$ and the upper bound for $t\geq t_{II}=5.16$, whereas the material purely made of phase 2 attains the lower bound for $t\geq  t_{I}=2.84$. For $t\leq t_{II}=5.16$ the upper bound is realized by a laminate%
\index{laminate}
 of the two components.}
\label{Hfig:Kel_r0}
\end{figure}
As shown in Fig. \ref{Hfig:Kel_r0}, the material purely made of phase 1 ($u_0=0$) attains the lower bound for $t\leq t_I=\frac{\eta_K}{G_K}\log\left(\frac{G_2}{G_2-G_K}\right)=2.78$ and the upper bound for $t\geq t_{II}=5.14$, whereas the material purely made of phase 2 ($u_0\to 1$) attains the lower bound for $t\geq t_I= 2.78$. For $t_I\leq t\leq t_{II}$ the upper bound is achieved by a laminate.
\begin{figure}[htbp]
\centering
\includegraphics[width=0.6\textwidth]{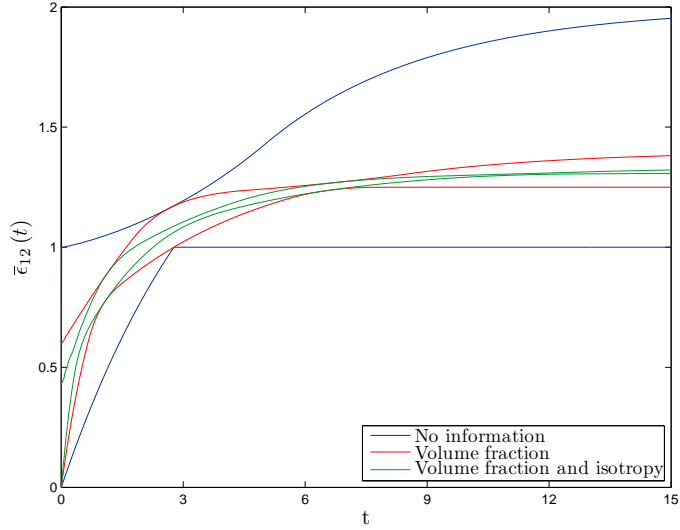}
\caption[The lower and upper bounds on $\overline{\epsilon}_{12}(t)$ with no information, volume fraction information, and with isotropy plus known volume fractions.]{Comparison between the lower and upper bounds on $\overline{\epsilon}_{12}(t)$ (normalized with respect to the elastic strain in phase 2, equal to $\sigma_0/(2G_2)$) in the following three cases: no information about the composite is given; the volume fraction of the components is known ($f_1=0.4$); and the composite is isotropic with given volume fractions.}
\label{Hfig:Kel_r0f1isotropy}
\end{figure}
\begin{figure}[htbp]
\centering
\includegraphics[width=0.6\textwidth]{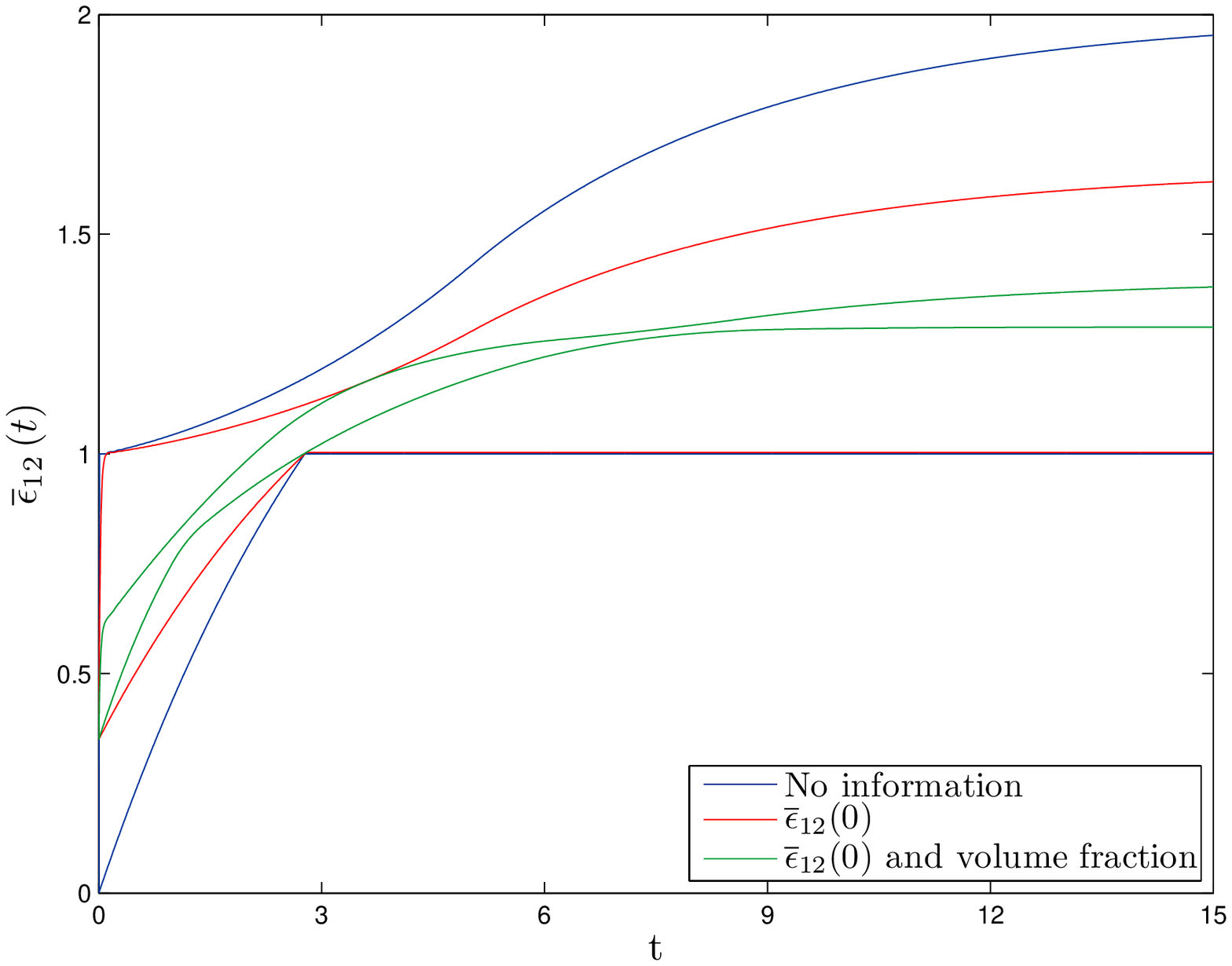}
\caption[The lower and upper bounds on $\overline{\epsilon}_{12}(t)$ with no information, known $\overline{\epsilon}_{12}(0)$, and known $\overline{\epsilon}_{12}(0)$ plus known volume fractions.]{Comparison between the lower and upper bounds on $\overline{\epsilon}_{12}(t)$ (normalized with respect to the elastic strain in phase 2, equal to $\sigma_0/(2G_2)$) in the following three cases: no information about the composite is given; the value of $\overline{\epsilon}_{12}(t)$ at $t=0$ is prescribed; and the value of $\overline{\epsilon}_{12}(t)$ at $t=0$ and the volume fractions are known ($f_1=0.4$). In the last two cases, the upper bound attains the assigned value of $\overline{\epsilon}_{12}(t)$ at $t=0$ only in the near vicinity of $t=0$.}
\label{Hfig:Kel_r0f1epsilon0}
\end{figure}
\begin{figure}[htbp]
\centering
\includegraphics[width=0.6\textwidth]{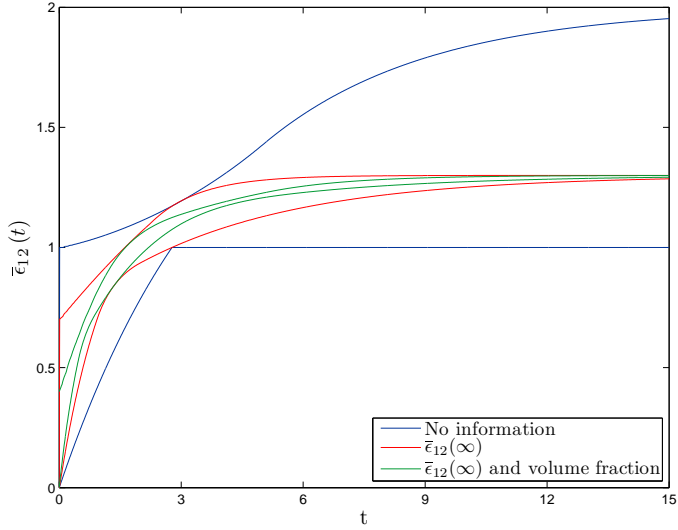}
\caption[The lower and upper bounds on $\overline{\epsilon}_{12}(t)$ with no information, known $\overline{\epsilon}_{12}(\infty)$, and known $\overline{\epsilon}_{12}(\infty)$ plus known volume fractions.]{Comparison between the lower and upper bounds on $\overline{\epsilon}_{12}(t)$ (normalized with respect to the elastic strain in phase 2, equal to $\sigma_0/(2G_2)$) in the following three cases: no information about the composite is given; the value of $\overline{\epsilon}_{12}(t)$ at $t\to\infty$ is prescribed; and the value of $\overline{\epsilon}_{12}(t)$ at $t\to\infty$ and the volume fractions are known ($f_1=0.4$). }
\label{Hfig:Kel_r0f1epsiloninf}
\end{figure}
Figs. \ref{Hfig:Kel_r0f1isotropy}, \ref{Hfig:Kel_r0f1epsilon0}, and \ref{Hfig:Kel_r0f1epsiloninf} depict the bounds on $\overline{\epsilon}_{12}(t)$ for different combinations of information about the composite. In particular, Fig.  \ref{Hfig:Kel_r0f1isotropy} shows the results when the volume fraction is known and the composite is transversely isotropic, Fig. \ref{Hfig:Kel_r0f1epsilon0} when $f_1$ and $\overline{\epsilon}_{12}(0)$ are assigned, and Fig. \ref{Hfig:Kel_r0f1epsiloninf} when $f_1$ and $\overline{\epsilon}_{12}(\infty)$ are prescribed. For each case, very tight bounds on $\overline{\epsilon}_{12}(t)$ are obtained.

With reference to Fig. \ref{Hfig:Kel_r0f1epsilon0}, it is worth noting that the upper bound attains the value $\overline{\epsilon}_{12}(0)$ by converging to such a value only in the near vicinity of $t=0$. This is due to the fact that the only non zero residue $P_{{11}}^{(0)}=(1-u_0)(1-G_2\overline{\epsilon}_{12}(0)/\sigma_0)$ tends to zero as $u_0\to 1$.\\
\\

One would like also to seek bounds on the possible values of the vector $\overline{\boldsymbol{\epsilon}}(t)=[\overline{\epsilon}_{12}(t)\,\,\, \overline{\epsilon}_{13}(t)]^T$%
\index{bounds!strain vector}
can take as time evolves. We do this by seeking bounds on a linear combination of the components $\overline{\epsilon}_{12}(t)$ and $\overline{\epsilon}_{13}(t)$ of $\overline{\boldsymbol{\epsilon}}(t)$. 
Let us consider, then, the following objective function, at a fixed angle $\alpha$:
 \begin{equation}\label{HG}
\mathcal{G}(t)=\sin\alpha\,\overline{\epsilon}_{12}(t)+\cos\alpha\,\overline{\epsilon}_{13}(t).
\end{equation}

We suppose that the following hypotheses still hold: phase 1 is described by the Kelvin-Voigt model,%
\index{Kelvin-Voigt model}
phase 2 has an elastic behavior, and the applied stress history is constant in time
for $t>0$. Then, equation \eqref{Hepsilon_con_trasformata} turns into 

\begin{equation}\label{Hepsilon_t}
\overline{\boldsymbol{\epsilon}}(t)=\frac{\boldsymbol{\sigma_0}}{2G_2}-\frac{1}{2G_2}\sum_{i=0}^m\left\{G_K-G_2+G_2\frac{\mathrm{exp}\left[-\frac{(G_K(1-u_i)+u_iG_2)t}{\eta_K\left(1-u_i\right)}\right]}{1-u_i}\right\}\frac{\mathbf{P}_i}{G_K-u_i(G_K-G_2)}\boldsymbol{\sigma}_0.
\end{equation}

We assume the composite has reflection symmetry and following the same argument adopted for deriving bounds on $\mathcal{F}(t)$ \eqref{HF}, we first fix the orientation of the composite
(i.e. residues), then for each time $t$ we minimize the function $\mathcal{G}(t)$ \eqref{HG}, where the components $\overline{\epsilon}_{13}(t)$ of $\overline{\boldsymbol{\epsilon}}(t)$ 
are given by \eqref{Hepsilon_t}, and we look for a function $\overline{\boldsymbol{\epsilon}}(t,\alpha)$ which achieves the minimum. Next, for each $t$ we construct the 
set which is the union of the points $\overline{\boldsymbol{\sigma}}(t,\alpha)$ as $\alpha$ varies between $0$ and $2\pi$, and take its convex hull. Finally, we take the union of the 
results as the orientation of the composite is varied.

In Videos 4 and 5, we plot the domain $\overline{\epsilon}_{13}(t)$-$\overline{\epsilon}_{12}(t)$ (where both strains have been normalized by the strain field in the elastic phase $\sigma_0/(2G_2)$) for each time $t\in[0,\infty)$, for the case when no information about the composite is available. In particular, in Video 4 we suppose one knows the orientation of the composite, while in Video 5 we suppose that such information is not available and, therefore, we consider the union of the domains calculated for each fixed orientation. Once again, the results are enriched by considering also the exact solution provided by a laminate. 

The optimum value of $\mathcal{G}(t)$ is attained when a maximum of two residues are non zero. At $t=0$, the strain field turns out to be $\overline{\boldsymbol{\epsilon}}(0)=(0,0)$ and, therefore, it does not depend on the position of the poles $u_0$ and $u_1$. For times $t>0$, instead, we maximize (or minimize) $\mathcal{G}(t)$ by varying the position of the two poles. The point $(0,1)$, corresponding to $u_0\to1,\,u_1\to 1$, keeps fixed since it represents the elastic response of phase 2. As the time goes by, the domain becomes smaller and smaller converging towards this point. At $t=t_I=2.78$, the domain coincides with the one representing the laminate response and, then, for $t>t_I=2.78$ it becomes bigger and bigger above the point $(0,1)$.

Assigned the angle $\alpha$ (see equation \eqref{HG}), bounds on $\mathcal{G}(t)$ are derived by considering the points of intersection between the domain $(\overline{\epsilon}_{13}(t),\overline{\epsilon}_{12}(t))$ and the tangents having slope equal to $\tan\alpha$, for each time $t$.

\section{Formulation of the problem}\label{HProblem}
\setcounter{equation}{0}

We consider a 3D body $\Omega$ made of a statistically homogeneous two-phase composite material with a
length scale of inhomogeneities much smaller than the length scale of the body (that is, $\Omega$ can be interpreted as the Representative Volume Element%
\index{Representative Volume Element}
of the composite), and subject on the boundary $\Gamma$ either to prescribed displacements or to assigned tractions, applied in such a way as to generate a \textit{shear antiplane} state within the solid. 

In case the volume average of the strain field, $\overline{\boldsymbol{\epsilon}}(t)$, is assigned, we choose kinematic boundary conditions%
\index{kinematic boundary conditions}
of the ``affine" type all over the surface $\Gamma$:
\begin{equation}\label{Hkinematic_boundary_conditions}
u_1(\mathbf{x},t)=2 H(t)\,\left(\overline{\epsilon}_{12}(t)\,x_2+\overline{\epsilon}_{13}(t)\,x_3\right),\quad u_2(\mathbf{x},t)=u_3(\mathbf{x},t)=0,
\end{equation}
with $H(t)$ the Heaviside unit-step function%
\index{Heaviside function}
of time, whereas in case the volume average of the stress field, $\overline{\boldsymbol{\sigma}}(t)$, is prescribed, we apply homogeneous tractions $\mathbf{p}(\mathbf{x},t)$ on $\Gamma$:
\begin{equation}\label{Hstatic_boundary_conditions}
p_1(\mathbf{x},t)=H(t)\,\left(\overline{\sigma}_{12}(t)\,n_2(\mathbf{x})+\overline{\sigma}_{13}(t)\,n_3(\mathbf{x})\right),\quad p_2(\mathbf{x},t)=p_3(\mathbf{x},t)=0,
\end{equation}
with $\mathbf{n}(\mathbf{x})$ the unit outward normal.

The local constitutive equations are given by \eqref{HConstitutive_law} and \eqref{HInverse_Constitutive_law}, while the effective constitutive laws are expressed by equation \eqref{HEffective_Constitutive_Laws}.

By applying the Laplace transform to \eqref{HEffective_Constitutive_Laws}, we obtain 
\begin{equation}\label{HComplex_effective_constitutive_law}
{\overline{\boldsymbol{\sigma}}}(\lambda)={\mathbf{C}}_*(\lambda){\overline{\boldsymbol{\epsilon}}}(\lambda), \quad\quad\quad
{\overline{\boldsymbol{\epsilon}}}(\lambda)={\mathbf{M}}_*(\lambda){\overline{\boldsymbol{\sigma}}}(\lambda),
\end{equation}
where the matrices $\mathbf{C}_*(\lambda)$ and $\mathbf{M}_*(\lambda)$ prove to be analytic functions of the eigenvalues $\mu_i(\lambda)$ and $\zeta_i(\lambda)$, $i=1,2$
(\citeAY{Bergman:1978:DCC}, \citeAY{Milton:1981:BCP}, \citeAY{Golden:1983:BEP}).
Consequently, by exploiting such analytic properties, an integral representation formula%
\index{integral representation formula}
for the operators $\mathbf{C}_*(\lambda)$ and $\mathbf{M}_*(\lambda)$ can be derived (for a rigorous mathematical proof, refer to the papers by Golden and Papanicolaou (\citeyearNP{Golden:1983:BEP}, \citeyearNP{Golden:1985:BEP})).

In particular, let us focus on the operator $\mathbf{C}_*(\lambda)$. By introducing the parameter $s(\lambda)$, defined by \eqref{Hs_u}, and the function $\mathbf{F}(s)$, given by \eqref{HF_G_definition}, 
\citeAPY{Golden:1983:BEP} enunciated and proved the so-called \textit{Representation theorem},%
\index{representation theorem}
which asserts that there exists a finite Borel measure%
\index{measure!matrix valued}
$\boldsymbol{\eta}(y)$, defined over the interval $[0,1]$ such that the measure is positive semi-definite matrix-valued satisfying
\begin{equation}\label{HF_Golden}
\mathbf{F}(s)= \int_0^1\frac{\mathrm{d}\boldsymbol{\eta}(y)}{s-y},
\end{equation}
for all $s\not\in[0,1]$. 
 
In the case when $\mathbf{C}_*(\lambda)$, and hence $\mathbf{F}(s)$, are rational functions the measure is concentrated at the poles $s_0,s_1,...,s_m$ of the rational function $\mathbf{F}(s)$ and equation \eqref{HF_Golden} turns into 
\begin{equation}\label{HF_rational}
\mathbf{F}(s)=\sum_{i=0}^m\frac{\mathbf{B}_{i}}{s-s_i},
\end{equation}
in which the poles%
\index{poles}
$s_i$ lie on the semi-closed interval $[0,1)$ and the residues%
\index{residue matrices}
$\mathbf{B}_i$ are positive semi-definite matrices, that is
\begin{equation}\label{HProperties_s_B}
0\leq s_0\leq s_1\leq ... \leq s_m<1\quad \mbox{and}\quad\mathbf{B}_i\geq 0 \,\,\,\mbox{for all}\,\, i.
\end{equation}

Notice that, since $\mathbf{C}_*(\lambda)$ is real and positive definite when the ratio $\mu_1(\lambda)/\mu_2(\lambda)$ is real and positive, and in particular as such a ratio tends to zero, from the definition \eqref{HF_G_definition} of $\mathbf{F}(s)$ it follows that, as $s\to 1$
\begin{equation}
\mathbf{F}(1)=\int_0^1\frac{\mathrm{d}\boldsymbol{\eta}(y)}{1-y}\leq \mathbf{I},
\end{equation}
and, in the case of rational functions, the latter reduces to the following constraint on the poles and residues of $\mathbf{F}(s)$:
\begin{equation}\label{Hconstraint0}
\mathbf{F}(1)=\sum_{i=0}^{m}\frac{\mathbf{B}_{i}}{1-s_i}\leq \mathbf{I}.
\end{equation}

In order to further reduce the number of free parameters $s_i$ and $\mathbf{B}_i$, all the available information about the composite microstructure has to be translated into constraints, the so-called \textit{sum rules},%
\index{sum rules}
on such parameters. In particular, the sum rules are obtained by expanding the representation \eqref{HF_Golden} of $\mathbf{F}(s)$ in powers of $1/s$ as $s\to\infty$, which corresponds to consider the case $\mu_1(\lambda)=\mu_2(\lambda)=1$, that is, when the microscopic structure is nearly homogeneous. When $s\to\infty$, the denominator in \eqref{HF_Golden} can be expanded as a series expansion%
\index{series expansion}
in powers of $1/s$ to give
\begin{equation}\label{HF_moments_measure}
\mathbf{F}(s)=\sum_{j=0}^{\infty}\frac{\mathbf{A}_j}{s^{j+1}}\quad\quad \mbox{with}\,\,\mathbf{A}_j=\int_0^1y^j\mathrm{d}\boldsymbol{\eta}(y).
\end{equation} 

It is clear that constraints on the moments%
\index{measure!moments}
of the measure are provided by the knowledge of the leading terms in the series, such as $\mathbf{A}_0$ and $\mathbf{A}_1$, which were derived through perturbation analysis%
\index{perturbation analysis}
by \citeAPY{Brown:1955:SMP} and \citeAPY{Bergman:1978:DCC}: see also equation (28) in \citeAPY{Milton:1981:BCP}. 
In particular, if the volume fractions $f_1$ and $f_2=1-f_1$ of the constituents are known, the first and second moments of the measure are given by
\begin{equation}
\mathbf{A}_0=\int_0^1\mathrm{d}\boldsymbol{\eta}(y)=f_1\mathbf{I},
\end{equation}
\begin{equation}
\mathrm{Tr}\mathbf{A}_1=\int_0^1y\,\mathrm{d}\boldsymbol{\eta}(y)=f_1\,f_2,
\end{equation} 
and the consequent constraints on the residues $\mathbf{B}_i$ and poles  $s_i$ read
\begin{equation}\label{Hconstraint_1}
\sum_{i=0}^m\mathbf{B}_i=f_1\mathbf{I},
\end{equation}
\begin{equation}\label{Hconstraint_2}
\mathrm{Tr}\left(\sum_{i=0}^m\mathbf{B}_is_i\right)=f_1\,f_2.
\end{equation} 
Concerning the inverse constitutive law operator $\mathbf{M}_*(\lambda)$, an analogous procedure leads to the following spectral representation:%
\index{representation formula}
\begin{equation}\label{HG_rational}
\mathbf{G}(u)= \sum_{i=0}^{m}\frac{\mathbf{P}_i}{u-u_i},
\end{equation}
where the parameter $u(\lambda)$ is defined by \eqref{Hs_u} and the function $\mathbf{G}(u)$ is given by \eqref{HF_G_definition}.

The residues $\mathbf{P}_i$ and poles $u_i$ satisfy the same constraints fulfilled by $\mathbf{B}_i$ and $s_i$. In particular, they satisfy inequalities \eqref{HProperties_s_B} and \eqref{Hconstraint0}, and equations \eqref{Hconstraint_1} and \eqref{Hconstraint_2}, provided one replaces $\mathbf{B}_i$ and $s_i$ with $\mathbf{P}_i$ and $u_i$.
 
\section{Sum rules}\label{HSum rules}
\setcounter{equation}{0}

The sum rules%
\index{sum rules}
we develop here are implicit in the work of \citeAPY{Bergman:1978:DCC}, but we reproduce them here for completeness. 
Let us consider the $\overline{\sigma}_{12}(t)$ component of the averaged stress field $\overline{\boldsymbol{\sigma}}(t)$ \eqref{Hsigma_con_trasformata}, given, in the most general case, by :
\begin{equation}\label{Hsigma12_con_trasformata}
\overline{\sigma}_{12}(t)=\mu_2(t)\ast\overline{\epsilon}_{12}(t)- \sum_{i=0}^m{B}_{11}^{(i)}\,\mathcal{L}^{-1}\left[\frac{\mu_2(\lambda)}{s-s_{i}}\right](t)\ast\overline{\epsilon}_{12}(t),
\end{equation}
where, for simplicity, we set $\overline{\epsilon}_{13}(t)=0$. In order to optimize the value of $\overline{\sigma}_{12}(t)$ for each $t\in[0,\infty)$ as a function of the $11$-components, ${B}_{11}^{(i)}$, of the residues $\mathbf{B}_i$, the constraints illustrated in Section \ref{HProblem} must be translated into constraints on ${B}_{11}^{(i)}$, non-negative quantities by virtue of \eqref{HProperties_s_B}. In particular, inequality \eqref{Hconstraint0}, rephrased as $\sum_{i=0}^m{\mathbf{e}^{\mathrm{T}}\mathbf{B}_i\mathbf{e}}/(1-s_i)\leq 1$, with $\mathbf{e}=[1\,\,\,0]^{\mathrm{T}}$, delivers 
\begin{equation}\label{Hconstraint0_bA}
1-\sum_{i=0}^m\frac{{B}_{11}^{(i)}}{1-s_i}\geq 0.
\end{equation}
We remark that given any set of poles $0\leq s_0 \leq s_1\leq s_2\leq\ldots \leq s_m<1$ and any set of non-negative residues ${B}_{11}^{(0)}, {B}_{11}^{(1)}, {B}_{11}^{(2)}, \ldots, {B}_{11}^{(m)}$ one can find a composite (which is a laminate of laminates)%
\index{laminate of laminates}
which realizes the response \eqref{Hsigma12_con_trasformata} for all times [see Appendix B of \citeAPY{Milton:1981:BTO} and Section 18.5 of \citeAPY{Milton:2002:TOC}]. This implies that all
our bounds based on the representation \eqref{Hsigma12_con_trasformata} will be optimal%
\index{bounds!optimality}
(and attained within this class of laminates of laminates), except those bounds that assume transverse
isotropy. The bounds assuming transverse isotropy will likely not be optimal as they fail to take into account the phase interchange relation%
\index{phase interchange relation}
of \citeAPY{Keller:1964:TCC}, 
which places a non-linear constraint on the residues.

By rephrasing the constraint \eqref{Hconstraint_1} as $\sum_{i=0}^m\mathbf{e}^{\mathrm{T}}\mathbf{B}_i\mathbf{e}=f_1$, we have
\begin{equation}\label{Hconstraint1_bA}
\sum_{i=0}^m{B}_{11}^{(i)}=f_1.
\end{equation}
Finally, by introducing the hypothesis of a transversely isotropic material (for which the residues $\mathbf{B}_i$ are diagonal matrices with ${B}_{11}^{(i)}={B}_{22}^{(i)}$), the constraint \eqref{Hconstraint_2} turns into
\begin{equation}\label{Hconstraint2_bA}
\sum_{i=0}^m{B}_{11}^{(i)}s_i=\frac{f_1f_2}{2}.
\end{equation}

Due to the linearity, with respect to ${B}_{11}^{(i)}$, of $\overline{\sigma}_{12}(t)$ and of the above constraints, we can apply the theory of linear programming%
\index{linear programming}
(\citeAY{Dantzig:1998:LPE}) to optimize $\overline{\sigma}_{12}(t)$, as shown in Section \ref{Hbounds}.\\ 
\\
In case the function to optimize is the scalar quantity $\mathcal{F}(t)$, defined by \eqref{HF}, the sum rules must be written in terms of the four components of the $2\times 2$ matrices $\mathbf{B}_i$. 

The constraint \eqref{HProperties_s_B} on the positive semi-definiteness of the residues $\mathbf{B}_i$ yields a condition on the determinant of $\mathbf{B}_i$, which is quadratic with respect to the components of $\mathbf{B}_i$. In order to have only linear constraints, we express the residues in the following form:
\begin{equation}\label{HB}
\mathbf{B}_i=\mathbf{R}^{\mathrm{T}}_i\mathbf{b}_i\mathbf{R}_i,\quad\quad i=0,1,...,m,
\end{equation}
with
\begin{equation}\label{Htheta_i}
\mathbf{R}_i=\left[
\begin{array}{cc}
\cos\theta_i&-\sin\theta_i\\
\sin\theta_i&\cos\theta_i
\end{array}\right],\quad
\quad\quad
\mathbf{b}_i=\left[
\begin{array}{cc}
b_{Ai}&0\\
0&b_{Bi}
\end{array}\right].
\end{equation}

Consequently, the condition on the positive semi-definiteness of the residues is translated into the following linear constraint on the elements $b_{Ai}$ and $b_{Bi}$, for $i=0,1,...,m$:
\begin{equation}\label{Hpositivity_residues}
b_{Ai}\geq 0 \quad\mbox{and}\quad b_{Bi}\geq 0.
\end{equation}
Regarding the constraint \eqref{Hconstraint0}, in order to avoid the condition of non-negativity of the determinant of the matrix $\mathbf{I}-\sum_{i=0}^m\mathbf{B}_i/(1-s_i)$, which is quadratic with respect to $b_{Ai}$ and $b_{Bi}$, we initially restrict our attention to the case of composites endued with \textit{reflective symmetry}.%
\index{reflective symmetry}
In such composites the angles of rotation $\theta_i$ \eqref{Htheta_i} take the same value for each residue, that is, the residues are diagonal matrixes with respect to the same basis, so that $\theta_i=\theta$ for every $i=0,1,...,m$, and the constraint \eqref{Hconstraint0} turns into the following linear conditions on $b_{Ai}$ and $b_{B_i}$:
\begin{equation}\label{Hconstraint0_reflective}
1-\sum_{i=0}^m\frac{b_{Ai}}{1-s_i}\geq 0,\quad\quad\quad 1-\sum_{i=0}^m\frac{b_{Bi}}{1-s_i}\geq 0.
\end{equation}
Furthermore, under the reflective symmetry property, relations \eqref{Hconstraint_1}, \eqref{Hconstraint_2} lead to
\begin{equation}\label{Hconstraint_1reflective}
\sum_{i=0}^mb_{Ai}=f_1,\quad\quad\quad\sum_{i=0}^mb_{Bi}=f_1f_2,
\end{equation}
\begin{equation}\label{Hconstraint_2reflective}
\sum_{i=0}^m(b_{Ai}+b_{Bi})s_i=f_1f_2.
\end{equation}
It is understood that in the case one would like to optimize the strain response, such as the $\overline{\epsilon}_{12}(t)$ component of the average stress field \eqref{Hepsilon_con_trasformata}:
\begin{equation}\label{Hepsilon12_con_trasformata}
\overline{\epsilon}_{12}(t)=\zeta_2(t)\ast\overline{\sigma}_{12}(t)- \sum_{i=0}^m{P}_{11}^{(i)}\,\mathcal{L}^{-1}\left[\frac{\zeta_2(\lambda)}{u-u_{i}}\right](t)\ast\overline{\sigma}_{12}(t),
\end{equation}
where we set $\overline{\sigma}_{13}(t)=0$, or the function $\mathcal{G}(t)$ \eqref{HG}, the constraints above still hold, provide we rephrase them in terms of the residues $\mathbf{P}_i$ and poles $u_i$ of the function $\mathbf{G}(u)$ \eqref{HF_G_definition}. Again, it is true that given any set of poles $0\leq u_0 \leq u_1 \leq u_2\leq\ldots \leq u_m<1$ 
and any set of non-negative residues ${P}_{11}^{(0)}, {P}_{11}^{(1)}, {P}_{11}^{(2)}, \ldots, {P}_{11}^{(m)}$ one can find a composite (which is a laminate of laminates)
which realizes the response \eqref{Hepsilon12_con_trasformata} for all times [see the last paragraph in Section 18.5 of \citeAPY{Milton:2002:TOC}]. This implies that all
our bounds based on the representation \eqref{Hepsilon12_con_trasformata} will be optimal (and attained within this class of laminates of laminates),%
\index{laminate of laminates}
 except those bounds that assume transverse isotropy.

\section{Derivation of bounds in the time domain}\label{Hbounds}
\setcounter{equation}{0}
The spectral representations \eqref{HF_rational} and \eqref{HG_rational} of the matrix valued functions $\mathbf{F}(s)$ and $\mathbf{G}(u)$, respectively, provide bounds on the response of the material expressed in terms of bounds on the stress component $\overline{\sigma}_{12}(t)$ \eqref{Hsigma12_con_trasformata} and on $\mathcal{F}(t)$ \eqref{HF} or on the strain component $\overline{\epsilon}_{12}(t)$ \eqref{Hepsilon12_con_trasformata} and on $\mathcal{G}(t)$ \eqref{HG}. These bounds are found by suitably varying the associated residues and poles in order to satisfy the sum rules shown in Section \ref{HSum rules}. Since the parameters $\mu_i(\lambda)$ and $\zeta_i(\lambda)$, $i=1,2$, are real it follows that $s(\lambda)$ and $u(\lambda)$ \eqref{Hs_u} are also real.

\subsection{Bounds on the stress response}\label{HBounds_stress}
\index{bounds!stress response}

By virtue of equations \eqref{HF_G_definition} and \eqref{HF_rational}, the direct complex effective constitutive law \eqref{HComplex_effective_constitutive_law} can then be rephrased as follows
\begin{equation}
\overline{\boldsymbol{\sigma}}(\lambda)=\mu_2(\lambda)\left[
\overline{\boldsymbol{\epsilon}}(\lambda)-\sum_{i=0}^m\frac{\mathbf{B}_{i}}{s-s_i}\overline{\boldsymbol{\epsilon}}(\lambda)\right],
\end{equation}
and by applying the inverse of the Laplace transform, the averaged stress field in the time domain is given by \eqref{Hsigma_con_trasformata}. Notice that in \eqref{Hsigma_con_trasformata} the inverse of the Laplace transform of ${\mu_2(\lambda)}/(s(\lambda)-s_{i})$ can be calculated explicitly, provided we know the functions $\mu_i(\lambda)$, $i=1,2$. 

Now the problem is to bound $\overline{\sigma}_{12}(t)$ \eqref{Hsigma12_con_trasformata} for each fixed value of $t$.
The idea is to take a fixed but large value of $m$ and find the maximum (or minimum) value of $\overline{\sigma}_{12}(t)$ as the poles $s_{i}$ and the non-negative components $B_{11}^{(i)}$ of the residues $\mathbf{B}_i$ are varied subject to the constraints \eqref{Hconstraint0_bA}, \eqref{Hconstraint1_bA} and \eqref{Hconstraint2_bA}. Since the resulting maximum (or minimum) could depend on $m$, we should ideally take the limit as $m$ tends to infinity. However, it turns out that the extremum does not depend on $m$, provided $m$ is large enough, and therefore there is no need to take limits. 

It is worth noting that varying the poles $s_i$ and the residues $\mathbf{B}_i$ corresponds, roughly speaking, to varying the microgeometry of the composite.
Therefore, the procedure described above may be compared to finding the maximum (or minimum) value of $\mathcal{F}(t)$ as the geometry of the composite is varied over all configurations. 
Strictly speaking this is not quite correct as not all combinations of poles $s_i$ and the residues $\mathbf{B}_i$ correspond to composites, as composites satisfy the phase interchange relation%
\index{phase interchange relation}
of  \citeAPY{Keller:1964:TCC}, which we have ignored as it places a non-linear constraint on the residues. This implies that the bounds we obtain assuming transverse isotropy, or the bounds we 
obtain by minimizing $\mathcal{F}(t)$ \eqref{HF} or $\mathcal{G}(t)$ \eqref{HG}, are probably not optimal (though we emphasize that our bounds on $\overline{\sigma}_{12}(t)$ and
$\overline{\epsilon}_{12}(t)$ which do not assume transverse isotropy are optimal).

\paragraph{No available information about the composite} In this case the maximum (or minimum) value of $\overline{\sigma}_{12}(t)$ is achieved when either one residue is non zero or all residues are zero. In particular, the extremum occurs  either when the constraint \eqref{Hconstraint0_bA} is satisfied as an equality by ${B}_{11}^{(0)}$, which takes the value ${B}_{11}^{(0)}=1-s_0$, 
while ${B}_{11}^{(i)}=0$, for $i=1,...,m$, or when ${B}_{11}^{(i)}=0$ for every $i=0,1,...,m$. Consequently, either
\begin{equation}\label{Hsigma12_no_info}
\overline{\sigma}_{12}(t)=\mu_2(t)\ast\overline{\epsilon}_{12}(t)- (1-s_0)\,\mathcal{L}^{-1}\left[\frac{\mu_2(\lambda)}{s(\lambda)-s_{0}}\right](t)\ast\overline{\epsilon}_{12}(t),
\end{equation}
with $s_0\in[0,1)$, or
\begin{equation}
\overline{\sigma}_{12}(t)=\mu_2(t)\ast\overline{\epsilon}_{12}(t).
\end{equation}
It is clear that the latter case is a subcase of \eqref{Hsigma12_no_info} when $s_0\to 1$, and corresponds to an isotropic material purely composed of phase 2, whereas when $s_0=0$ in \eqref{Hsigma12_no_info}, by means of the definition \eqref{Hs_u} of $s(\lambda)$, \eqref{Hsigma12_no_info} provides the stress state in an isotropic material purely composed of phase 1, i.e., $\overline{\sigma}_{12}(t)=\mu_1(t)\ast\overline{\epsilon}_{12}(t)$. All that remains (and in general this is best done numerically) is to find, for each time $t$, 
the position of the pole $s_0$ which maximizes or minimizes \eqref{Hsigma12_no_info}.

The upper and lower limits of the function \eqref{Hsigma12_no_info} are given by equations \eqref{Hsigma_max} and \eqref{Hsigma_min} and they are shown in Fig. \ref{Hfig:Max_noinfo} for the specific case when the response of one phase is given by the Maxwell model%
\index{Maxwell model}
and the other having purely elastic behavior, with constant applied strain history.

\paragraph{The volume fraction of the constituents is known} If $f_1$ is prescribed, then $\overline{\sigma}_{12}(t)$ is optimized by considering either only one non zero residue satisfying constraint \eqref{Hconstraint1_bA} or only two non zero residues fulfilling the constraint \eqref{Hconstraint1_bA} and relation \eqref{Hconstraint0_bA} as an equality. In the first case, ${B}_{11}^{(0)}=f_1$ and
\begin{equation}\label{Hsigma12_volume_fractions_one_pole}
\overline{\sigma}_{12}(t)=\mu_2(t)\ast\overline{\epsilon}_{12}(t)- f_1\,\mathcal{L}^{-1}\left[\frac{\mu_2(\lambda)}{s(\lambda)-s_{0}}\right](t)\ast\overline{\epsilon}_{12}(t),
\end{equation}
with $s_0\in[0,f_2]$, whereas in the second case ${B}_{11}^{(0)}=\frac{(1-s_0)(s_1-f_2)}{s_1-s_0}$, ${B}_{11}^{(1)}=\frac{(1-s_1)(f_2-s_0)}{s_1-s_0}$ and 
\begin{equation}\label{Hsigma12_volume_fractions}
\begin{split}
\overline{\sigma}_{12}(t)=\mu_2(t)\ast\overline{\epsilon}_{12}(t)&- \frac{(1-s_0)(s_1-f_2)}{s_1-s_0}\,\mathcal{L}^{-1}\left[\frac{\mu_2(\lambda)}{s(\lambda)-s_{0}}\right](t)\ast\overline{\epsilon}_{12}(t)\\&-\frac{(1-s_1)(f_2-s_0)}{s_1-s_0}\,\mathcal{L}^{-1}\left[\frac{\mu_2(\lambda)}{s(\lambda)-s_{1}}\right](t)\ast\overline{\epsilon}_{12}(t),
\end{split}
\end{equation}
with $s_0\in[0,f_2]$ and $s_1\in[f_2,1)$.

We point out that equation \eqref{Hsigma12_volume_fractions_one_pole} is a specific case of \eqref{Hsigma12_volume_fractions}, when the pole $s_1$ approaches 1.
The remaining optimization over the position of the poles in general needs to be done numerically.

Fig. \ref{Hfig:Max_r0f1isotropy} shows the bounds obtained from equation \eqref{Hsigma12_volume_fractions}, in case phase 1 is modeled by the Maxwell model and phase 2 has an elastic behavior, 
with the further assumption that the strain history is constant.

\paragraph{The composite is isotropic with known volume fractions} Bounds on $\overline{\sigma}_{12}(t)$ can then be derived by either considering two non zero residues satisfying equations \eqref{Hconstraint1_bA} and \eqref{Hconstraint2_bA}, so that ${B}_{11}^{(0)}=f_1\,\frac{s_1-f_2/2}{s_1-s_0}$, ${B}_{11}^{(1)}=f_1\,\frac{f_2/2-s_0}{s_1-s_0}$ (subject to the constraint that the inequality
\eqref{Hconstraint0_bA} is satisfied) or by taking only three residues to be non zero, with \eqref{Hconstraint0_bA} holding as an equality, so that
\begin{gather}\label{Hsigma_12isotropy}
{B}_{11}^{(0)}=\frac{(1-s_0)(1-s_1)(1-s_2)}{(s_1-s_0)(s_2-s_0)}\left[1-\frac{f_1}{1-s_2}-f_1\,\frac{s_2-f_2/2}{(1-s_1)(1-s_2)}\right],\\\nonumber {B}_{11}^{(1)}=\frac{(1-s_0)(1-s_1)(1-s_2)}{(s_1-s_0)(s_2-s_1)}\left[\frac{f_1}{1-s_0}+f_1\,\frac{f_2/2-s_0}{(1-s_0)(1-s_2)}-1\right],\\\nonumber {B}_{11}^{(2)}=\frac{(1-s_0)(1-s_1)(1-s_2)}{(s_2-s_0)(s_2-s_1)}\left[1-\frac{f_1}{1-s_0}-f_1\,\frac{f_2/2-s_0}{(1-s_0)(1-s_1)}\right].
\end{gather}
Again the remaining optimization over the position of the poles in general needs to be done numerically.
This case is shown in Fig. \ref{Hfig:Max_r0f1isotropy} for the Maxwell model-Elastic model case with constant strain history.\\
\\
Apart from the knowledge of the volume fractions and of the possible isotropy of the composite, other information may be given. For instance, the value of $\overline{\sigma}_{12}(t)$ at $t=0$ or at $t\to \infty$ may be known. In such a case, we can derive bounds on $\overline{\sigma}_{12}(t)$ as follows:
\paragraph{Given value of $\overline{\sigma}_{12}(t)$ at $t=0$ or at $t\to \infty$} The maximum (or minimum) value of the 12-component of the averaged stress field can be obtained either by considering only one non zero residue satisfying equation \eqref{Hsigma12_con_trasformata} evaluated at $t=0$ or at $t\to\infty$, respectively, or only two non zero residues fulfilling constraint \eqref{Hsigma12_con_trasformata} (evaluated at $t=0$) and relation \eqref{Hconstraint0_bA} as an equality. \\
\\
It is worth noting that tighter bounds can be derived by considering combinations of information, such as the value of $\overline{\sigma}_{12}(t)$ at zero or infinity and the volume fraction of the material (see Figs.\ref{Hfig:Max_r0f1sigma0} and \ref{Hfig:Max_r0f1sigmainf}). For the sake of brevity we do not report here the explicit results for that case but it is understood that they are derived following the same procedure applied above.  \\
\\
Now let us look at the problem of bounding the function $\mathcal{F}(t)$ \eqref{HF} for a composite with reflective symmetry,%
\index{reflective symmetry}
with the angles $\alpha$ and $\theta$ fixed. 
%
%
%
%

\paragraph{Bounds in case no information about the composite is available} 

In case the only available information about the composite is the shear modulus $\mu_i(\lambda)$ of each constituent, then bounds on $\mathcal{F}(t)$ \eqref{HF} have to be sought by considering the constraints \eqref{Hpositivity_residues} and \eqref{Hconstraint0_reflective}. The optimum value of $\mathcal{F}(t)$ is attained when maximum two residues are non zero. In particular, the representative case can be considered as the one for which both the constraints given by \eqref{Hconstraint0_reflective} are fulfilled as equalities. Then, only one of the $b_{Ai}$ elements and only one of the $b_{Bj}$ elements, with $i\neq j$, are non zero, that is, either $b_{A0}=1-s_0$ and $b_{B1}=1-s_1$ or $b_{A1}=1-s_1$ and $b_{B0}=1-s_0$, where $s_0$ has to be varied over $[0,1)$ and $s_1$ over $[s_0,1)$ to give the optimum value of $\mathcal{F}(t)$. Note that the second case can be recovered from the first one, by switching the angle $\theta$ to $\theta+\pi/2$ (see equation \eqref{Htheta_i}). Let us consider, then, the first option. The corresponding expression for the averaged stress field $\overline{\boldsymbol{\sigma}}(t)$ \eqref{Hsigma_con_trasformata} reads:
\begin{equation}\label{Hsigma_no_info_2poles}
\begin{split}
\overline{\boldsymbol{\sigma}}(t)=\mu_2(t)\ast\overline{\boldsymbol{\epsilon}}(t)&- (1-s_0)
\left[ 
\begin{array}{cc}
\cos^2\theta&-\sin\theta\,\cos\theta\\
-\sin\theta\,\cos\theta&\sin^2\theta
\end{array}\right]\,\mathcal{L}^{-1}\left[\frac{\mu_2(\lambda)}{s-s_0}\right](t)\ast\overline{\boldsymbol{\epsilon}}(t)\\&
- (1-s_1)
\left[ 
\begin{array}{cc}
\sin^2\theta&\sin\theta\,\cos\theta\\
\sin\theta\,\cos\theta&\cos^2\theta
\end{array}\right]\,\mathcal{L}^{-1}\left[\frac{\mu_2(\lambda)}{s-s_1}\right](t)\ast\overline{\boldsymbol{\epsilon}}(t),
\end{split}
\end{equation}
and the maximum (or minimum) value of $\mathcal{F}(t)$ has to be determined by varying the poles $s_0$ and $s_1$ over the respective validity intervals. Finally the union of the resulting possible
values of $\overline{\boldsymbol{\sigma}}(t)$ is taken as $\theta$ is varied (see Video 3). This case can be considered as the representative combination because, when either the poles approach 1
(with the associated residue tending to zero) or take the same value, all the other possible combinations can be derived consequently. 

\paragraph{Bounds in case the volume fractions are known} In case the volume fractions $f_1$ and $f_2$ of the constituents are known, bounds on $\mathcal{F}(t)$ \eqref{HF} can be derived by considering also the constraints provided by equations \eqref{Hconstraint_1reflective} and \eqref{Hconstraint_2reflective}. Specifically, the maximum (or minimum) value of the function $\mathcal{F}(t)$ is attained by one of the combinations which range from the two poles case to the five poles case. In the former situation, the bound is realized by considering either two non zero $b_{Ai}$ and one non zero $b_{Bj}$, where $j$ is equal to one of the two $i$, or vice versa. In the five poles case, instead, the bound on $\mathcal{F}(t)$ is attained by considering those $b_{Ai}$ and $b_{Bj}$ which satisfy \eqref{Hconstraint_1reflective}-\eqref{Hconstraint_2reflective} and constraints \eqref{Hconstraint0_reflective} as equalities, that is, by considering either three non zero $b_{Ai}$ and two non zero $b_{Bj}$, with $i\neq j$, or vice versa. We stress the fact that the five poles case is the representative one (and the only one which needs to be considered) in the sense that all the other combinations can be consequently recovered by letting some poles collapse to the same value or approach 1. \\
\\

\subsection{Bounds on the strain response}\label{HBounds_strain}%
\index{bounds!strain response}

Let us consider the complex effective inverse constitutive law \eqref{HComplex_effective_constitutive_law}. Thanks to the relation between $\mathbf{M}_*(\lambda)$ and $\mathbf{G}(u)$, given by \eqref{HF_G_definition}, and the spectral representation \eqref{HG_rational} of the function $\mathbf{G}(u)$, the averaged strain field in the complex domain is then described by the following equation: 
\begin{equation}
\overline{\boldsymbol{\epsilon}}(\lambda)=\zeta_2(\lambda)\left[
\overline{\boldsymbol{\sigma}}(\lambda)- \sum_{i=0}^m\frac{\mathbf{P}_{i}}{u-u_i}\overline{\boldsymbol{\sigma}}(\lambda)\right],
\end{equation}
while in the time domain, by applying the inverse of the Laplace transform, $\overline{\boldsymbol{\epsilon}}(t)$ is given by
\eqref{Hepsilon_con_trasformata}.

In this case, the problem consists in bounding the $\overline{\epsilon}_{12}(t)$ component \eqref{Hepsilon12_con_trasformata} of the averaged strain field. Alternatively, the aim could be the optimization of the function $\mathcal{G}(t)$ \eqref{HG}. 
In both cases, following the same arguments adopted in Subsection \ref{HBounds_stress}, bounds analogous to those obtained for $\mathcal{F}(t)$ and  $\overline{\sigma}_{12}(t)$ can be deduced also for $\mathcal{G}(t)$ and $\overline{\epsilon}_{12}(t)$, respectively. 

\section{Composites without reflective symmetry}
\labsect{HGeneralization}
\setcounter{equation}{0}

Bounds on the functions $\mathcal{F}(t)$ \eqref{HF} and $\mathcal{G}(t)$ \eqref{HG} have been derived under the hypothesis of reflective symmetry. In particular, such an assumption allows one to derive linear constraints on the diagonal elements $b_{Ai}$ and $b_{Bi}$ of the matrixes $\mathbf{b}_i$ \eqref{Htheta_i}. Nevertheless, in the case the composite is not symmetric with respect to a certain plane, that is, the reflective symmetry assumption does not hold, we can still derive linear constraints on the elements $b_{Ai}$ and $b_{Bi}$. 

To see this, let us introduce an additional pole $s_{m+1}=1-\delta$, where $\delta$ is a sufficiently small parameter, with residue 
$$
\mathbf{B}_{m+1}={\delta}{\mathbf{D}},\quad\quad
\mathbf{D}=\mathbf{I}-\sum_{i=0}^{m} \frac{\mathbf{B}_i}{1-s_i}. 
$$
Then, the introduction of a fictitious pole%
\index{fictitious pole}
with very small residue does not affect the bounds on the analytic function, except in the near vicinity of $s=1$.
Consequently, inequality \eqref{Hconstraint0} can be replaced by the following equality:
\begin{equation}\label{Hconstraint0_new}
\sum_{i=0}^{m+1}\frac{\mathbf{B}_i}{1-s_i}=\mathbf{I},
\end{equation} 
which provides three linear constraints with respect to the $b_{Ai}$ and $b_{Bi}$:

\begin{gather}\label{Hconstraint_0new}\nonumber
\sum_{i=0}^{m+1}\frac{b_{Ai}\cos^2\theta_i+b_{Bi}\sin^2\theta_i}{1-s_i}=1, \quad \quad
\sum_{i=0}^{m+1}\frac{b_{Ai}\sin^2\theta_i+b_{Bi}\cos^2\theta_i}{1-s_i}=1,\\
\sum_{i=0}^{m+1}\frac{(b_{Ai}-b_{Bi})\cos\theta_i\sin\theta_i}{1-s_i}=0.
\end{gather}
Finally, relations \eqref{Hconstraint_1} and \eqref{Hconstraint_2} written in terms of the $b_{Ai}$ and $b_{Bi}$ lead, respectively, to
\begin{gather}\label{Hconstraint_1new}
\sum_{i=0}^{m}{b_{Ai}\cos^2\theta_i+b_{Bi}\sin^2\theta_i}=f_1 ,\quad \quad
\sum_{i=0}^{m}{b_{Ai}\sin^2\theta_i+b_{Bi}\cos^2\theta_i}=f_1,\\
\sum_{i=0}^{m}{(b_{Ai}-b_{Bi})\cos\theta_i\sin\theta_i}=0,
\end{gather}
and
\begin{equation}\label{Hconstraint_2new}
\sum_{i=0}^m\left(b_{Ai}+b_{Bi}\right)s_i=f_1f_2.
\end{equation}

In contrast to the case with reflective symmetry,
the bounds on $\mathcal{F}(t)$ (as $\alpha$ is varied), for fixed $t$, necessarily restrict $\overline{\boldsymbol{\sigma}}(t)$ to a convex region in the $(\overline{\sigma}_{12}(t), \overline{\sigma}_{13}(t))$ plane. However the range of values of $\overline{\boldsymbol{\sigma}}(t)$, as the poles and residue matrices are varied (subject to the constraints \eqref{Hconstraint_0new},
and, if the volume fractions are known, \eqref{Hconstraint_1new} and \eqref{Hconstraint_2new}) is in fact a convex set in the $(\overline{\sigma}_{12}(t), \overline{\sigma}_{13}(t))$ plane.
To see this, suppose $m$ is enormously large. Then there is no loss of generality if we take the poles to be evenly spaced: $s_i=i/(m+2)$, and take the angles $\theta_i$ to 
increase by small amounts going in total many times ``around the clock'': $\theta_i=2\pi(m\,\,{\rm mod}\,\,k)/k$, where $k$ is chosen with $m\gg k \gg 1$, and only vary the $b_{Ai}$ and $b_{Bi}$. 
Then, if a set of parameters $b_{Ai}$ and $b_{Bi}$, $i=0,1,\ldots, m$ satisfy the constraints, and another set $b_{Ai}'$ and $b_{Bi}'$ also satisfy it, so will the linear combination
$wb_{Ai}+(1-w)b_{Ai}'$ and $wb_{Ai}+(1-w)b_{Ai}'$, for any weight $w\in (0,1)$ and the resulting response vector $\overline{\boldsymbol{\sigma}}_w(t)$ will be a linear combination of the two
response vectors, $\overline{\boldsymbol{\sigma}}(t)$ and $\overline{\boldsymbol{\sigma}}'(t)$ associated with the original two sets of parameters.

In the following, we show the procedure to be adopted in order to derive bounds on the function $\mathcal{F}(t)$ \eqref{HF}. In contrast to the case with reflective symmetry,
the bounds on $\mathcal{F}(t)$ (as $\alpha$ is varied) for fixed $t$ necessarily restrict $\overline{\boldsymbol{\sigma}}(t)$ to a convex region in the $(\overline{\sigma}_{13}(t), \overline{\sigma}_{12}(t))$ plane. Another method needs to be devised to obtain bounds that confine $\overline{\boldsymbol{\sigma}}(t)$ to regions that 
are not-necessarily convex in the $(\overline{\sigma}_{13}(t), \overline{\sigma}_{12}(t))$ plane.

\paragraph{Bounds in the case where no information about the composite is available} For the sake of brevity, we do not report the explicit expression taken by the stress field \eqref{Hsigma_con_trasformata} for each combination of poles related to this case but we consider only the representative case. In particular, the optimal value of $\mathcal{F}(t)$ is attained when either only one or only three residues are non zero. In particular, the representative combination of residues corresponds to the case for which the three constraints given by \eqref{Hconstraint0_new} are fulfilled. Such a condition holds when either only three elements among the $b_{Ai}$ are non zero, while $b_{Bi}=0$ for every $i=0,1,...,m$, and vice versa, or when only two elements among the $b_{Ai}$ and one element among the $b_{Bj}$, with $i\neq j$, are non zero, and vice versa. It is worth noting that, by suitably choosing the angles $\theta_i$ \eqref{Htheta_i}, the latter case is equivalent to the former one.

\paragraph{Bounds in the case when the volume fractions are known} The combinations of residues which provide the maximum (or minimum) value of $\mathcal{F}(t)$ are those which satisfy the seven equations given by the constraints \eqref{Hconstraint0_new}, \eqref{Hconstraint_1new} and \eqref{Hconstraint_2new}. In particular, the combination with the minimum number of poles is given by three non zero $b_{Ai}$ and the corresponding three non zero $b_{Bi}$ (three poles in total), while the combination with the maximum number of poles consists of seven poles and can be achieved either considering six non zero $b_{Ai}$ and one non zero $b_{Bj}$, $i\neq j$, and vice versa, or five non zero $b_{Ai}$ and two non zero $b_{Bj}$, $i\neq j$, and vice versa, or four non zero $b_{Ai}$ and three non zero $b_{Bj}$, $i\neq j$, and vice versa. We remark that all combinations corresponding to the same number of poles are equivalent, since we are free to replace each rotation angle $\theta_i$ \eqref{Htheta_i} by $\theta_i+\pi/2$. We emphasize that the seven pole case is the representative one (and the only one which needs to be considered) in the sense that all the other combinations can be consequently recovered by letting some poles collapse to the same value or approach 1 (implying that the associated residue tends to zero). 
\section{Bounding the homogenized relaxation and creep kernels}
\index{bounds!relaxation kernel}%
\index{bounds!creep kernel}
\setcounter{equation}{0}
Note that the relation \eqref{Hsigma_con_trasformata} when $\overline{\boldsymbol{\epsilon}}(t)$ is chosen to be a constant $\boldsymbol{\epsilon}_0$ for all $t>0$ can be written in the form
\begin{equation}\label{Hrelax}
\overline{\boldsymbol{\sigma}}(t)=\boldsymbol{C}^h(t)\boldsymbol{\epsilon}_0,
\end{equation}
where $\boldsymbol{C}^h(t)$, the homogenized relaxation kernel,%
\index{homogenized relaxation kernel}
is given by
\begin{equation}\label{Hrelaxh}
\boldsymbol{C}^h(t)=\mu_2(t)- \sum_{i=0}^m\mathbf{B}_{i}\,\mathcal{L}^{-1}\left[\frac{\mu_2(\lambda)}{s(\lambda)-s_{i}}\right](t).
\end{equation}
The same arguments that were used in the previous section to show
that the range of values of $\overline{\boldsymbol{\sigma}}(t)$, as the poles and residue matrices are varied is in fact a convex set, can also be applied here:
the range of values of the matrix valued relaxation kernel $\boldsymbol{C}^h(t)$ 
 as the poles and residue matrices are varied (subject to any linear sum rules on the residues, implied by the  known information
about the composite) is also a convex set. 

To find this convex set we consider for each fixed time $t$ the objective function%
\index{objective function}
\begin{equation}
\mathcal{F}({\bf V})=\mathrm{Tr}({\bf V}\boldsymbol{C}^h(t)),
\end{equation}
where ${\bf V}$ is any $2\times 2$ real valued symmetric matrix. By substituting \eqref{Hrelaxh} in this expression we see that the objective function depends linearly on the residue matrices%
\index{residue matrices}
$\mathbf{B}_{i}$, and thus we can use the same
techniques as before to find the minimum values of $\mathcal{F}$ for a given matrix ${\bf V}$ (incorporating, if desired, known information
about the composite which impose sum rules on the residues): let us call this minimum
$\mathcal{F}^{\min}({\bf V})$. The constraint that 
\begin{equation}
\mathrm{Tr}({\bf V}\boldsymbol{C}^h(t))\geq \mathcal{F}^{\min}({\bf V})
\end{equation}
confines $\boldsymbol{C}^h(t)$ to lie on one side of a ``hyperplane'' in a $3$-dimensional space with the elements of $\boldsymbol{C}^h(t)$ as coordinates (as it is a symmetric $2\times 2$ matrix
there are only $3$ independent elements). Finally, by varying ${\bf V}$ we constrain $\boldsymbol{C}^h(t)$ to the desired convex set in this 3-dimensional space.

In a similar way the relation \eqref{Hepsilon_con_trasformata} when $\overline{\boldsymbol{\sigma}}(t)$ is chosen to be a constant, $\boldsymbol{\sigma}_0$, for all $t>0$ can be written in the form
\begin{equation}
\overline{\boldsymbol{\epsilon}}(t)=\boldsymbol{M}^h(t)\boldsymbol{\sigma}_0,
\end{equation}
where $\boldsymbol{M}^h(t)$, the homogenized creep kernel,%
\index{homogenized creep kernel}
is given by
\begin{equation}\label{Hcreeph}
\boldsymbol{M}^h(t)=\zeta_2(t)- \sum_{i=0}^m\mathbf{P}_{i}\,\mathcal{L}^{-1}\left[\frac{\zeta_2(\lambda)}{u-u_{i}}\right](t).
\end{equation}
As $\boldsymbol{M}^h(t)$ depends linearly on the residues $\mathbf{P}_{i}$ we can also use the same approach to bound it (subject to any linear sum rules on the residues, 
implied by the  known information about the composite).
\section{Correlating the transient response to different applied fields at different times}
\setcounter{equation}{0}
We have been focusing on deriving bounds on the transient response of the composite at a single time $t$, and for a single applied field. However, if desired, the method allows one to obtain coupled bounds%
\index{bounds!coupled}
which correlate the responses at a set of different times $t=t_1, t_2,\ldots, t_n$, and for different applied fields (which may or may not be all the same). To see this, suppose for example that 
we are interested in coupling the stresses $\overline{\boldsymbol{\sigma}}^{(j)}(t^{(j)})$, for $j=1,2,\ldots,n$ that arise respectively 
in response to the applied strains $\overline{\boldsymbol{\epsilon}}^{(j)}(t)$, for $j=1,2,\ldots,n$. From \eqref{Hsigma_con_trasformata} it directly follows that
\begin{equation}\label{Hsigma_con_trasformataj}
\overline{\boldsymbol{\sigma}^{(j)}}(t^{(j)})=\mu_2(t^{(j)})\ast\overline{\boldsymbol{\epsilon}}(t^{(j)})
- \sum_{i=0}^m\mathbf{B}_{i}\,\mathcal{L}^{-1}\left[\frac{\mu_2(\lambda)}{s-s_{i}}\right](t^{(j)})\ast\overline{\boldsymbol{\epsilon}}^{(j)}(t^{(j)}).
\end{equation}
 The same arguments that were used in Section \sect{HGeneralization} to show
that the range of values of $\overline{\boldsymbol{\sigma}}(t)$, as the poles and residue matrices are varied is in fact a convex set, can also be applied here:
the range of values of the $n$-tuple $(\overline{\boldsymbol{\sigma}}^{(1)}(t^{(1)}), \overline{\boldsymbol{\sigma}}^{(2)}(t^{(2)}), \ldots, \overline{\boldsymbol{\sigma}}^{(n)}(t^{(n)}))$
 as the poles and residue matrices are varied (subject to any linear sum rules on the residues, implied by the  known information
about the composite) is also a convex set.

To find this convex set,  consider the objective function%
\index{objective function}
\begin{equation}\label{HFobject}
\mathcal{F}({\bf v}^{(1)},{\bf v}^{(2)},\ldots,{\bf v}^{(n)})=\sum_{j=1}^{n}{\bf v}^{(j)}\cdot\boldsymbol{\sigma}^{(j)}(t^{(j)}).
\end{equation}
By substituting \eqref{Hsigma_con_trasformataj} in this expression we see that the objective function depends linearly on the residue matrices $\mathbf{B}_{i}$, and thus we can use the same
techniques as before to find the minimum values of $\mathcal{F}$ for a given set of vectors ${\bf v}^{(1)},{\bf v}^{(2)},\ldots,{\bf v}^{(n)}$ (incorporating, if desired, known information
about the composite which impose sum rules on the residues): let us call this minimum
$\mathcal{F}^{\min}({\bf v}^{(1)},{\bf v}^{(2)},\ldots,{\bf v}^{(n)})$. The constraint that 
\begin{equation}\label{HFconst}
\sum_{j=1}^{n}{\bf v}^{(j)}\cdot\boldsymbol{\sigma}^{(j)}(t^{(j)})\geq \mathcal{F}^{\min}({\bf v}^{(1)},{\bf v}^{(2)},\ldots,{\bf v}^{(n)})
\end{equation}
confines the $n$-tuple $(\overline{\boldsymbol{\sigma}}^{(1)}(t^{(1)}), \overline{\boldsymbol{\sigma}}^{(2)}(t^{(2)}), \ldots, \overline{\boldsymbol{\sigma}}^{(n)}(t^{(n)}))$ to lie on
one side of a ``hyperplane'' in a $2n$-dimensional space with the elements of the $\overline{\boldsymbol{\sigma}}^{(j)}(t^{(j)})$ as coordinates. Finally by varying the vectors
${\bf v}^{(1)}$, ${\bf v}^{(2)}$, $\ldots$, ${\bf v}^{(n)}$ we constrain the $n$-tuple to the desired convex set in this multidimensional space. 

Note that the applied strains 
$\overline{\boldsymbol{\epsilon}}^{(j)}(t)$ could all be identical, and in this case the bounds will correlate the values of the resulting stress field
$\overline{\boldsymbol{\sigma}}(t)$ at times $t=t_1, t_2,\ldots, t_n$.  These bounds, correlating the transient response to different applied fields at a set of different times,
might be very useful for predicting the response to a new applied field, given measurements (at specific times) for the response to a set of test applied fields. Or they could be very
useful if used in an inverse fashion to determine information about the composite, such as the volume fractions of the phases. 

It is clear that the method can easily be extended in the obvious way to obtain bounds which correlate the matrix values of the relaxation kernel%
\index{bounds!relaxation kernel}
$\boldsymbol{C}^h(t)$ \eqref{Hrelaxh} at different times or the creep kernel%
\index{bounds!creep kernel}
$\boldsymbol{M}^h(t)$ \eqref{Hcreeph} at different times.

\section{Concluding remarks}\label{HConclusions}
\setcounter{equation}{0}

In this investigation, which constitutes a chapter of the book \textit{Extending the Theory of Composites to
Other Areas of Science} edited by G.W. Milton, we proposed a new approach to derive bounds on the response of a two-component viscoelastic composite under antiplane loadings, in the time domain. The starting point is represented by the so-called analytic method,%
\index{analytic method}
first proposed by \citeAPY{Bergman:1978:DCC} to bound effective conductivities when the component conductivities are real, and later extended to 
bound the complex effective tensor of a two-component dielectric composite in the frequency domain (see, for instance, Milton (\citeyearNP{Milton:1980:BCD}, \citeyearNP{Milton:1981:BCP}, \citeyearNP{Milton:1981:BTO}), and 
Bergman (\citeyearNP{Bergman:1980:ESM})) but, to the best of our knowledge, the method until now has been applied only in the frequency domain, for cyclic external actions at a certain frequency. This work may be the first 
to extend the field of applicability of the analytic method to problems defined in the time domain with non-cyclic external actions. 

The core of the analytic method is based on the fact that, by virtue of the analyticity property of the complex effective tensor of the viscoelastic composite with respect to the complex moduli of the components, one can write the complex effective tensor as the sum of poles weighted by positive semi-definite matrix valued residues. Consequently, the response of the material, in terms of stresses or strains, turns out to depend only on the position of the poles and on the value of the associated residues, which are the variational parameters of the problem. The aim is to find the combinations of such parameters which provide the maximum (or minimum) response of the composite for each moment of time. 

The optimization of the response of the material is performed in two steps. First, all the available information about the composite, such as the knowledge of the volume fraction of the constituents or of the value of the response of the material at a certain moment of time, is translated into (linear) constraints on the poles and residues. Then, the response of the material being linear in the residues, allows one to apply the theory of linear programming to limit the number of non zero residues, so that the problem is reduced to a new one with a relatively small number of non zero residues. Finally, the optimization over the positions of the poles is performed numerically for two specific cases: when the stress response has to be bounded, we consider a composite made of an elastic phase and a phase with a behavior describable by the Maxwell model,%
\index{Maxwell model}
whereas when we bound the strain response, we consider a composite made of an elastic phase and a phase modeled by the Kelvin-Voigt model.%
\index{Kelvin-Voigt model}

The estimates given by the numerical results prove to be increasingly accurate the more information about the composite is incorporated. In particular, when information such as the volume fraction of the components or the value of the response at a specific time is considered, the bounds are quite tight over the entire range of time, thus allowing one to predict the transient behavior of the composite. Most noticeably, when combinations of information are considered, such as the knowledge of the volume fractions and the eventual transverse isotropy of the composite, the bounds are extremely tight at certain specific times, suggesting the possibility of measuring the response of such times and, by using the bounds in an inverse fashion, almost exactly determining the volume fraction of the components of the composite. 

\section*{Acknowledgments}
Ornella Mattei is grateful for support from the Italian Ministry of Education, University, and Research (MIUR), from the University of Brescia, and from the University of Utah. Graeme Milton is grateful to the American National Science Foundation (Research Grant DMS-1211359) and the University of Utah for support.


\bibliography{tcbook,newref}
\bibliographystyle{mod-xchicago}

\end{document}